\documentclass[12pt]{article}
\usepackage{graphics}
\usepackage{epsfig}
\newcommand{\gsim}       {\mbox{\raisebox{-0.4ex}{$\;\stackrel{>}{\scriptstyle \sim}\;$}}}
\newcommand{\lsim}       {\mbox{\raisebox{-0.4ex}{$\;\stackrel{<}{\scriptstyle \sim}\;$}}}
\newcommand {\DR} {N_{T}+\epsilon N_{L}}
\newcommand {\eps} {\epsilon}
\newcommand {\R} {{\rm Re}}
\newcommand {\I} {{\rm Im}}
\begin{document}                                                    
                    
\title {
\LARGE \bf{
Measurement of the Spin-Density Matrix Elements in Exclusive Electroproduction
of \boldmath{$\rho^0$} Mesons at HERA}
}
\author{ZEUS Collaboration}

\date{}

\maketitle
\begin{abstract}
Exclusive electroproduction of $\rho^0$ mesons has been measured using the
ZEUS detector at HERA in two $Q^2$ ranges, $0.25<Q^2<0.85$~GeV$^2$ 
and $3<Q^2<30$~GeV$^2$. The low-$Q^2$ data span the range 
$20<W<90$~GeV; the high-$Q^2$ data cover the
$40<W<120$~GeV interval. Both samples extend up to four-momentum transfers
 of $|t|=0.6$~GeV$^2$. The distribution in the azimuthal
angle between the positron scattering plane and the $\rho^0$
production plane shows a small but significant violation of $s$-channel 
helicity conservation, corresponding to the production
of longitudinally polarised (i.e. helicity zero) $\rho^0$ mesons from 
transverse photons.
Measurements of the 15 combinations of spin-density matrix elements
which completely define the angular 
distributions are presented and discussed.
\end{abstract}

\vspace{-14.5cm}
\begin{flushleft}
\tt DESY 99-102 \\
July 1999 \\
\end{flushleft}

\thispagestyle{empty}
\newpage

%
%
%
%
\topmargin-1.cm                                                                                    
\evensidemargin-0.3cm                                                                              
\oddsidemargin-0.3cm                                                                               
\textwidth 16.cm                                                                                   
\textheight 680pt                                                                                  
\parindent0.cm                                                                                     
\parskip0.3cm plus0.05cm minus0.05cm                                                               
\def\3{\ss}                                                                                        
\newcommand{\address}{ }                                                                           
\pagenumbering{Roman}                                                                              
                                                   %
\begin{center}                                                                                     
{                      \Large  The ZEUS Collaboration              }                               
\end{center}                                                                                       
  J.~Breitweg,                                                                                     
  S.~Chekanov,                                                                                     
  M.~Derrick,                                                                                      
  D.~Krakauer,                                                                                     
  S.~Magill,                                                                                       
  B.~Musgrave,                                                                                     
  A.~Pellegrino,                                                                                   
  J.~Repond,                                                                                       
  R.~Stanek,                                                                                       
  R.~Yoshida\\                                                                                     
 {\it Argonne National Laboratory, Argonne, IL, USA}~$^{p}$
% ...
\newpage                                                                                           
% ...
$^{\    2}$ also at IROE Florence, Italy \\                                                        
% ...
$^{  35}$ partially supported by the Foundation for German-Russian Collaboration                   
DFG-RFBR \\ \hspace*{3.5mm} (grant no. 436 RUS 113/248/3 and no. 436 RUS 113/248/2)\\              
% ...
                                                           %
                                                           %
\newpage   
                                                           %
                                                           %
\begin{tabular}[h]{rp{14cm}}                                                                       
$^{a}$ &  supported by the Natural Sciences and Engineering Research                               
          Council of Canada (NSERC)  \\                                                            
$^{b}$ &  supported by the FCAR of Qu\'ebec, Canada  \\                                            
$^{c}$ &  supported by the German Federal Ministry for Education and                               
          Science, Research and Technology (BMBF), under contract                                  
          numbers 057BN19P, 057FR19P, 057HH19P, 057HH29P, 057SI75I \\                              
$^{d}$ &  supported by the MINERVA Gesellschaft f\"ur Forschung GmbH, the                          
German Israeli Foundation, and by the Israel Ministry of Science \\                                
$^{e}$ &  supported by the German-Israeli Foundation, the Israel Science                           
          Foundation, the U.S.-Israel Binational Science Foundation, and by                        
          the Israel Ministry of Science \\                                                        
$^{f}$ &  supported by the Italian National Institute for Nuclear Physics                          
          (INFN) \\                                                                                
$^{g}$ &  supported by the Japanese Ministry of Education, Science and                             
          Culture (the Monbusho) and its grants for Scientific Research \\                         
$^{h}$ &  supported by the Korean Ministry of Education and Korea Science                          
          and Engineering Foundation  \\                                                           
$^{i}$ &  supported by the Netherlands Foundation for Research on                                  
          Matter (FOM) \\                                                                          
$^{j}$ &  supported by the Polish State Committee for Scientific Research,                         
          grant No. 115/E-343/SPUB/P03/154/98, 2P03B03216, 2P03B04616,                             
          2P03B10412, 2P03B03517, and by the German Federal                                        
          Ministry of Education and Science, Research and Technology (BMBF) \\                     
$^{k}$ &  supported by the Polish State Committee for Scientific                                   
          Research (grant No. 2P03B08614 and 2P03B06116) \\                                        
$^{l}$ &  partially supported by the German Federal Ministry for                                   
          Education and Science, Research and Technology (BMBF)  \\                                
$^{m}$ &  supported by the Fund for Fundamental Research of Russian Ministry                       
          for Science and Edu\-cation and by the German Federal Ministry for                       
          Education and Science, Research and Technology (BMBF) \\                                 
$^{n}$ &  supported by the Spanish Ministry of Education                                           
          and Science through funds provided by CICYT \\                                           
$^{o}$ &  supported by the Particle Physics and                                                    
          Astronomy Research Council \\                                                            
$^{p}$ &  supported by the US Department of Energy \\                                              
$^{q}$ &  supported by the US National Science Foundation                                          
\end{tabular}                                                                                      
                                                           %
                                                           %

\newpage

\topmargin-1.5cm
\evensidemargin-0.3cm
\oddsidemargin-0.3cm
\textwidth 16.cm
\textheight 650pt
\parindent0.cm
\parskip0.3cm plus0.05cm minus0.05cm

\pagenumbering{arabic} 
\setcounter{page}{1}

\section{Introduction}

Exclusive production of vector mesons by real and virtual photons, 
$\gamma p\rightarrow Vp$ and  $\gamma^* p\rightarrow Vp$ 
($V=\rho^0, \omega, \phi, J/\psi, \dots$), has been 
studied extensively, both in fixed-target experiments and at HERA~\cite{jim}.
For photon-proton ($\gamma^*p$) centre-of-mass energies $W \gsim 10$~GeV,
the production of light vector mesons ($\rho^0, \omega, \phi$) 
at low photon virtuality, $Q^2 < 1$~GeV$^2$, 
exhibits features typical of soft diffractive processes: 
a weak dependence on centre-of-mass energy and a differential cross section 
that, at low $|t|$ values,  falls exponentially with $-t$, where $t$ is 
the square of the 
four-momentum exchanged between the photon and the proton. These features are 
consistent with the expectations of the Vector Meson Dominance model 
(VMD)~\cite{sakurai} according to which the photon 
is considered to fluctuate into a vector meson which then scatters elastically 
from the proton.

The low-energy data indicate that the amplitude for the photon/vector-meson 
transition is predominantly $s$-channel helicity conserving, i.e. the helicity 
of the vector meson is equal to that of the photon when 
the spin-quantisation axis is chosen along the direction of the meson momentum 
in the $\gamma^*p$ centre-of-mass system. 
The exchange of an object in the $t$-channel with $P=(-1)^J$, 
e.g. $J^P=0^+, 1^-, 2^+, 3^-,\dots$(natural parity exchange 
in the $t$ channel, NPE), dominates
such diffractive processes~\cite{bauer,collins}.
However, small helicity-single-flip and 
helicity-double-flip contributions to the amplitude have been reported 
in $\pi^+\pi^-$ photoproduction in the $\rho^0$ mass region at 
$W \lsim 4$~GeV~\cite{ballam}.
Helicity-single-flip amplitudes have also been observed in $\rho^0$ 
electroproduction for $1.3<W<2.8$~GeV and $0.3<Q^2<1.4$~GeV$^2$~\cite{joos}.
A helicity-single-flip contribution of $(14\pm 8)\%$ was measured 
in $\rho^0$ muoproduction at $W=17$~GeV~\cite{chio}.

Data from HERA have shown that when the reaction involves a large scale, such
as the charm quark mass in $J/\psi$ photoproduction~\cite{psi} or high $Q^2$ 
in $\rho^0$ electroproduction~\cite{rhodis95,h1}, the cross section rises 
more steeply with 
$W$ and the $t$ distribution is flatter than in the case where no such scale
is present, such as $\rho^0$ 
photoproduction~\cite{zeusrho93,h1rho93,zeusrho94}. Such results
can be understood in terms of models based on perturbative QCD (pQCD).
In these models, the photon fluctuates into a $q \bar{q}$ pair and 
the interaction of the pair with the proton is mediated by the exchange of 
two gluons in a colour singlet state~\cite{workshop96}. The resulting
cross section is proportional to the square of the gluon density in the proton.

The cross section for the exclusive production of vector mesons 
from virtual photons has contributions
from both transverse and longitudinal photons.  These contributions are
expected to have different $W$ and $Q^2$ dependences~\cite{jim,cfs}, and their
extraction would provide an important test of the 
understanding of the production process.  
The angular distributions of the decay products of the
vector meson, in principle, yield the separate contributions of 
transverse and longitudinal photons. 
Previous studies at 
HERA~\cite{rhodis95,zeusrho93,h1rho93,zeusrho94,zeusrhodis93,h1rhodis93,h1rhodiss94}
assumed conservation of helicity
in the $s$-channel amplitudes ($s$-channel helicity conservation, SCHC), 
in which case the ratio
of longitudinal to transverse photon 
cross sections, $R$, is given in terms of the ratio of the number of helicity 0 
(i.e. longitudinally polarised) to helicity $\pm 1$ 
(i.e. transversely polarised) 
$\rho^0$ mesons produced.  Sufficient data are now available to 
test the validity of SCHC at HERA by measuring the full set of matrix
elements which completely determine the decay angular distributions.
This paper reports the determination of such a set of matrix elements for
exclusive $\rho^0$ electroproduction
at HERA, $ep \rightarrow e \rho^0p$ ($\rho^0 \rightarrow \pi^+ \pi^-$).
Two kinematic ranges are studied, $0.25<Q^2<0.85$~GeV$^2$, $20<W<90$~GeV 
(referred to as the ``BPC" sample) and 
$3<Q^2<30$~GeV$^2$, $40<W<120$~GeV (the ``DIS" sample). 
The two samples are those already used in the previous ZEUS
paper~\cite{rhodis95}, where SCHC was assumed. A similar analysis has recently
been presented by the H1 Collaboration in the range 
$1<Q^2<60$~GeV$^2$~\cite{h1}.

The data are used to determine $R$ without relying on the SCHC
hypothesis, to evaluate the size of the helicity-single-flip and
helicity-double-flip amplitudes and to test the extent to which natural 
parity exchange dominates in the $t$ channel. 
The results are compared to recent 
calculations~\cite{ivanov,kolya,cudell} of the amplitudes for exclusive 
$\rho^0$ meson production in the framework of pQCD models of the type 
described above, i.e. assuming the $t$-channel exchange of a gluon pair. 

Under the assumption of NPE, there are five independent combinations of 
photon and meson helicity states, yielding 
two helicity-conserving amplitudes, two single-flip amplitudes
and one double-flip amplitude for $\rho^0$ electroproduction 
(see, for example,~\cite{ivanov}).
The calculations indicate that, for photon virtualities exceeding the hadronic 
mass scale of about 1~GeV$^2$, the single-flip amplitude for 
producing longitudinally polarised vector mesons from transverse photons is
significant. This amplitude is of higher twist and vanishes
in the non-relativistic limit of the $\rho^0$ meson wave function;
in addition, it is proportional to $\sqrt{|t|}$, while
the double-flip amplitude grows linearly with $|t|$. These 
amplitudes are rendered experimentally accessible via 
their interference with the dominant helicity-conserving amplitudes. 
Conversely, the amplitude for producing 
transversely polarised $\rho^0$ mesons from longitudinal photons is
expected to be negligible~\cite{diehl}.
There are also predictions of non-vanishing single and 
double-flip contributions in the non-perturbative region of 
small photon virtualities, $Q^2~<~1$~GeV$^2$~\cite{kolya,cudell}. 
VMD-based models~\cite{schildknecht} also predict
SCHC breaking on the basis of the analogy between photon-hadron and
hadron-hadron interactions. 

\section{Experimental set-up}

The data were collected at the $ep$ collider HERA in 1995 using the ZEUS 
detector.
In this period HERA operated at a proton energy of 820~GeV and a positron
energy of 27.5~GeV, giving a total centre-of-mass energy 
$\sqrt{s} \simeq 300$~GeV. 
A detailed description of the ZEUS detector
can be found elsewhere~\cite{rhodis95,zeus}.
The main components used in this analysis are briefly described below.

The high-resolution uranium-scintillator calorimeter, CAL,
consists of three parts:  forward~\footnote{Throughout this paper the 
standard ZEUS right-handed
coordinate system is used: the $Z$-axis points in the direction of
the proton beam momentum (referred to as the 
forward direction) and the horizontal $X$-axis
points towards the centre of HERA. The nominal interaction
point is at $X=Y=Z=0$.}
(FCAL),
barrel  (BCAL) and  rear (RCAL) calorimeters.
Each part is subdivided transversely into towers that
are segmented longitudinally  into one electromagnetic
section and one (RCAL) or two (FCAL, BCAL)
hadronic sections.

Charged-particle tracks  are reconstructed and their momenta measured
using  the central tracking detector (CTD). 
The CTD is a cylindrical drift chamber operated 
in a magnetic field of 1.43 T
produced by a  superconducting solenoid.
It consists of 72 cylindrical layers, organised in 9 superlayers
covering the polar angular region $15^\circ  < \theta < 164^\circ $.

The trajectory of positrons scattered at small angles with respect to the
beam direction is determined from the beam pipe calorimeter (BPC)
and the small-angle rear tracking detector (SRTD). 
The BPC is an electromagnetic sampling calorimeter located at $Z=-294$~cm.
The~SRTD~is attached to the front face of the RCAL.
It consists of two planes of scintillator strips, 1 cm wide
and 0.5 cm thick, arranged in orthogonal orientations  and read out
via optical fibres and photomultiplier tubes. It covers the region
of $68\times68$ cm$^2$ in $X$ and $Y$ with the  exclusion of a
$10\times20$ cm$^2$ hole at the centre for the beam pipe. 

\section{Kinematics and decay angular distributions}
\label{kinematics}

\begin{figure}[t] 
\begin{center}
\epsfig{figure=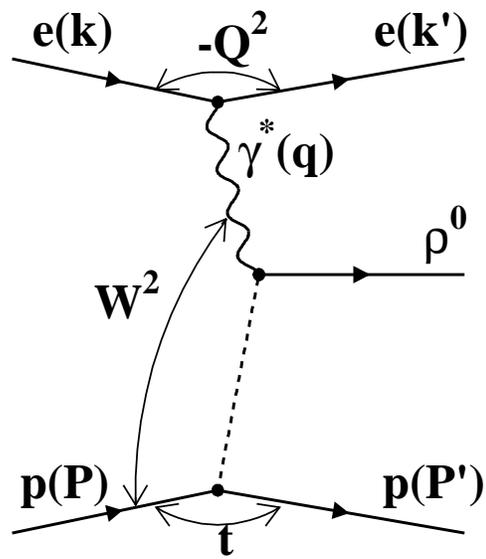,height=15cm}
\caption{Schematic diagram of the reaction $ep \rightarrow e \rho^0 p$,
indicating the kinematic variables used in this analysis.}
\label{diagram} 
\end{center} 
\end{figure}

The following kinematic variables, some of which are indicated in 
Fig.~\ref{diagram},
are used to describe exclusive $\rho^0$ production:
\begin{itemize}
\item The four-momenta of the incident positron ($k$), scattered positron 
($k^{\prime}$), incident proton ($P$), 
scattered proton ($P^{\prime}$) and virtual 
photon ($q$);
\item $Q^2=-q^2=-(k-k^{\prime})^2$, the negative  four-momentum squared of 
the virtual photon; 
\item $W^2 = (q+P)^2$, the squared centre-of-mass energy of the 
photon-proton system; 
\item $y = (P\cdot q)/(P\cdot k)$, the 
fraction of the positron energy transferred to the proton 
in its rest frame; 
\item $M_{\pi\pi}$, the invariant mass of the two decay pions;
\item $t = (P-P^{\prime})^2$,  the squared four-momentum transfer
at the proton vertex. 
\end{itemize}

The kinematic variables were  reconstructed using 
the so-called ``constrained'' method~\cite{rhodis95,rhodis94}, which uses
the momenta of the decay  particles measured in  the CTD and
the polar and azimuthal angles of the scattered positron determined with 
the  BPC (for the BPC sample) or with the CAL and the SRTD (for the DIS sample).

\begin{figure}[t] 
\begin{center}
\epsfig{figure=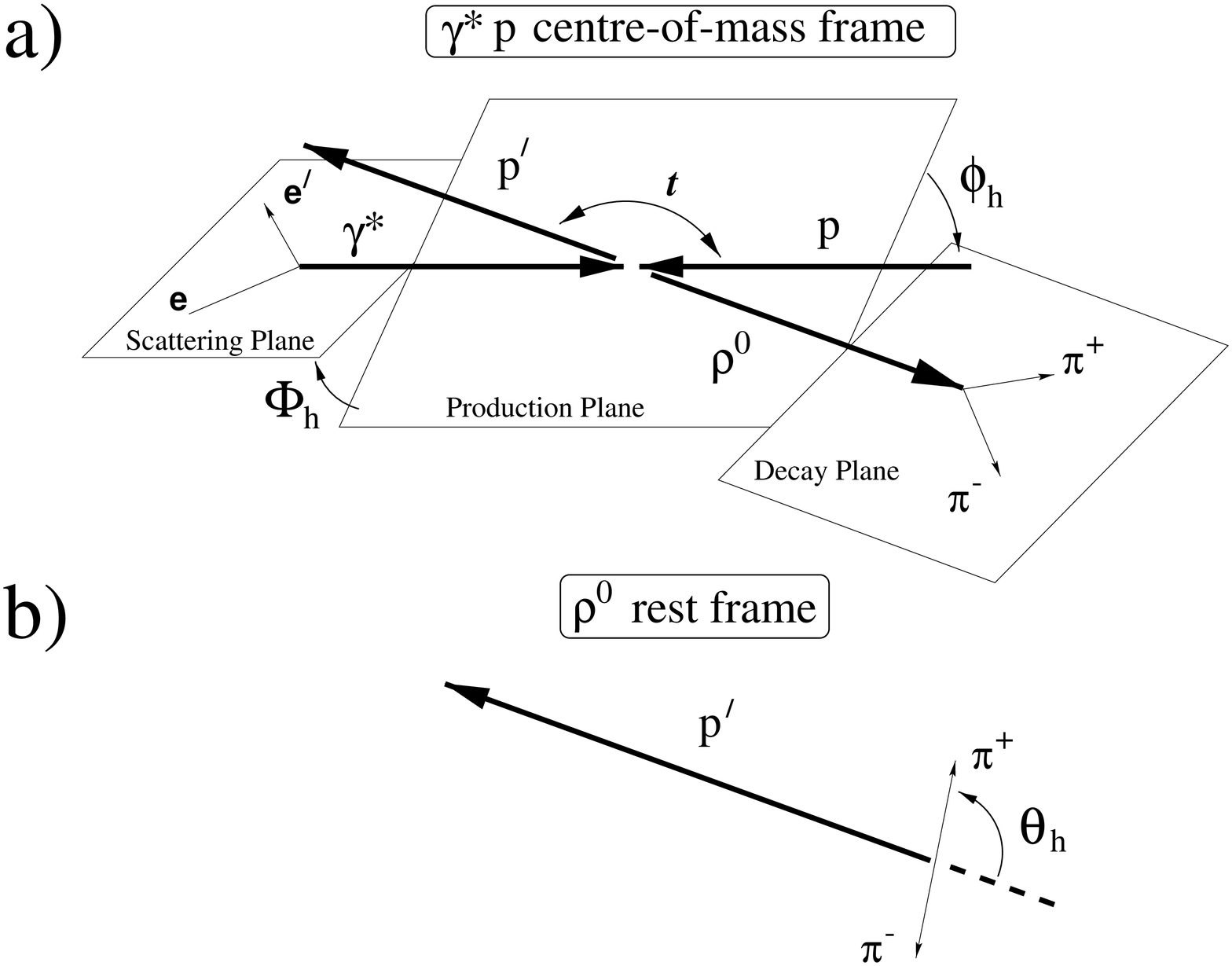,height=9cm}
\caption{
Schematic diagrams of (a) the process $ep \rightarrow e \rho^0 p$ in the
$\gamma^* p$ centre-of-mass system, and (b) the decay of the
$\rho^0$ in its rest frame. Three angles suffice to describe the reaction: 
the azimuthal angle between the scattering plane and the production plane, 
$\Phi_h$; and the two $\rho^0$ decay angles, $\phi_h$, the azimuthal angle 
between the production and decay planes, defined in either 
the $\gamma^* p$ system or in
the $\rho^0$ rest frame; and $\theta_h$, which is the polar angle of the 
positively-charged decay product defined with respect to the direction
of the $\rho^0$ momentum vector in the $\gamma^* p$ system, or, 
equivalently, the direction 
opposite to the momentum-vector of the final-state
proton in the rest frame of the $\rho^0$ meson. This choice of the
spin-quantisation axis defines the helicity frame.
}
\label{angles} 
\end{center} 
\end{figure}

The exclusive electroproduction and decay of $\rho^0$ mesons is
described, at given values of $W$, $Q^2$, $M_{\pi\pi}$ and $t$, by three angles:
$\Phi_h$ --  the angle between the $\rho^0$ production plane and the
positron scattering plane in the $\gamma^* p$ centre-of-mass frame (see
Fig.~\ref{angles});
$\theta_{h}$ and $\phi_{h}$ --  the polar and azimuthal angles 
of the positively charged decay pion in the $s$-channel helicity
frame, in which  the spin-quantisation  axis is defined 
as the direction opposite to the momentum of the final-state proton
in the $\rho^0$ rest frame.
In both the $\rho^0$ rest frame and the $\gamma^* p$ centre-of-mass
system, $\phi_{h}$ is the angle between the decay plane 
and the $\rho^0$ production plane.
The angular distribution as a function  of these three angles, 
$W(\cos\theta_{h},\phi_{h},\Phi_{h})$, is parameterised by the 
$\rho^0$ spin-density matrix elements, $\rho_{ik}^{\alpha}$,
where $i,k=-1,0,1$ and  by convention $\alpha$=0,1,2,4,5,6 for 
an unpolarised charged-lepton beam~\cite{ref:angle}. 
The superscript denotes the decomposition of the
spin-density matrix into contributions from the following photon polarisation
states: unpolarised transverse photons (0); linearly polarised
transverse photons (1,2); longitudinally polarised photons (4); and
from the interference of the longitudinal and transverse amplitudes (5,6). 

The decay angular distribution can be expressed in terms of
combinations, $r^{04}_{ik}$
and $r^{\alpha}_{ik}$, of the density matrix elements:
\begin{eqnarray}
r^{04}_{ik} &=& {\rho^0_{ik} \, + \, \epsilon R \rho^{4}_{ik}\over 1 \,
+ \, \epsilon R}, \label{r04}\\
r^{\alpha}_{ik} &=& \,
\left\{
\begin{array}{ll}
{\displaystyle {\rho^{\alpha}_{ik}\over 1 \, + \, \epsilon R}}, &
{\alpha}=1,2\\*[5mm]
{\displaystyle {\sqrt{R} \; \rho^{\alpha}_{ik}\over 1 \, + \, \epsilon
R}}, & {\alpha}=5,6,
\end{array} \right.
\end{eqnarray}

\noindent
where $\epsilon$ is 
the ratio of the longitudinal to transverse photon fluxes and 
$R=\sigma_{L}^{\gamma^{\star}{p}}/\sigma_{T}^{\gamma^{\star}{p}}$, with
$\sigma_{L}^{\gamma^{\star}{p}}$ and $\sigma_{T}^{\gamma^{\star}{p}}$
the cross sections for exclusive $\rho^0$ production from longitudinal
and transverse virtual photons, respectively.
In the kinematic 
range of this analysis, the value of $\epsilon$ varies only 
between 0.98 and 1.0; hence $\rho^{0}_{ik}$ and $\rho^{4}_{ik}$  
cannot be distinguished.

The hermitian nature of the spin-density matrix and the requirement
of parity conservation reduces the number of
independent combinations to 15, in terms of which
the angular distribution can be written as
\begin{eqnarray}
W(\cos{\theta_h},\phi_h,\Phi_h) &= & \frac{3}{4\pi} \biggl[ 
\frac{1}{2}(1-r^{04}_{00})+\frac{1}{2}(3r^{04}_{00}-1)\cos^2{\theta_h} \nonumber\\
& &-\sqrt{2}~{\rm Re}\{r^{04}_{10}\}\sin{2\theta_h}\cos{\phi_h} -r^{04}_{1-1}\sin^2{\theta_h}\cos{2\phi_h} \nonumber\\
& &-\epsilon\cos{2\Phi_h}(r^1_{11}\sin^2{\theta_h}+r^1_{00}\cos^2{\theta_h}-\sqrt{2}~{\rm Re}\{r^1_{10}\}\sin{2\theta_h}\cos{\phi_h}\nonumber\\
& &\hspace*{2.1cm}-r^1_{1-1}\sin^2{\theta_h}\cos{2\phi_h}) \nonumber\\
& &-\epsilon\sin{2\Phi_h}(\sqrt{2}~{\rm Im}\{r^2_{10}\}\sin{2\theta_h}\sin{\phi_h}+{\rm Im}\{r^2_{1-1}\}\sin^2{\theta_h}\sin{2\phi_h})\nonumber\\
& & +\sqrt{2\epsilon (1+\epsilon)}\cos{\Phi_h} (r^5_{11}\sin^2{\theta_h}+r^5_{00}\cos^2{\theta_h} \nonumber\\
& &\hspace*{3.7cm}-\sqrt{2}~{\rm Re}\{r^5_{10}\}\sin{2\theta_h}\cos{\phi_h} -r^5_{1-1}\sin^2{\theta_h}\cos{2\phi_h})\nonumber\\
& & +\sqrt{2\epsilon (1+\epsilon)}\sin{\Phi_h} (\sqrt{2}~{\rm Im}\{r^6_{10}\}\sin{2\theta_h}\sin{\phi_h} \nonumber\\
& &\hspace*{3.7cm}+{\rm Im}\{r^6_{1-1}\}\sin^2{\theta_h}\sin{2\phi_h}) \biggr].
\label{full_equation}
\end{eqnarray}

The 15 coefficients  $r^{04}_{ik}$, $r^{\alpha}_{ik}$ are
related to various combinations of 
the helicity amplitudes $T_{\lambda_{\rho} \lambda_{\gamma}}$, where 
$\lambda_{\rho}$ and $\lambda_{\gamma}$ are the helicities of the $\rho^0$
meson and of the photon, respectively. Table~\ref{amplitudes} shows the
relations between the  coefficients $r^{04}_{ik}$, $r^{\alpha}_{ik}$
and the helicity amplitudes.

Under the SCHC assumption, the 
angular distribution for the decay of the $\rho^0$ meson 
depends on only two angles, $\theta_{h}$ 
and $\psi_{h}=\phi_{h}-\Phi_{h}$, and is characterised
by three independent parameters, $r^{04}_{00}, r^1_{1-1}$ and Re$\{r^5_{10}\}$:
\begin{eqnarray}\label{eq:angular}
W(\cos{\theta_{h}},\psi_{h})&  = & \frac{3}{4\pi}
[\frac{1}{2}(1-r^{04}_{00})+\frac{1}{2}(3r^{04}_{00}-1) \
\cos^2{\theta_{h}} \nonumber  \\
& & +\;\epsilon r^1_{1-1} \sin^2{\theta_{h}} \cos{2\psi_{h}} \
-2\sqrt{\epsilon(1+\epsilon)} {\rm Re}\{r^5_{10}\} \sin{2\theta_{h}}
\cos{\psi_{h}}]. 
\end{eqnarray}
The SCHC hypothesis also implies that $r_{1-1}^1=-{\rm Im}\{r_{1-1}^2\}$ and 
${\rm Re}\{r_{10}^5\}=-{\rm Im}\{r_{10}^6\}$. In this case,
the ratio $R$ can be determined from the polar angular distribution alone:
\begin{equation}\label{eq:R_ratio}
R = \frac{1}{\epsilon}  \frac{r^{04}_{00}}{1 - r^{04}_{00}}.
\end{equation}
\noindent

The combined assumptions of SCHC and NPE further reduce the number of 
independent parameters to two, since the polar and azimuthal angular 
distributions are then related via:   
\begin{equation}\label{eq:NPE}
1-r^{04}_{00}-2r^1_{1-1}=0. 
\end{equation}

If the SCHC requirement is relaxed, then
\begin{equation}\label{eq:R_ratio_no_schc}
R = \frac{1}{\epsilon} \frac{r^{04}_{00}-\Delta^2}{1 - (r^{04}_{00}-\Delta^2)},
\end{equation}
\noindent
with
\begin{equation}\label{eq:delta2}
\Delta^2 = \frac{|T_{01}|^2 - 2\epsilon |T_{10}|^2}{N_T+\epsilon N_L},
\end{equation}
\noindent
where $N_L=|T_{00}|^2+2|T_{10}|^2$ and $N_T=|T_{11}|^2+|T_{1-1}|^2+|T_{01}|^2$. The
quantity $\Delta$ can be approximated as:
\begin{equation}\label{eq:delta}
\Delta \simeq \frac{|T_{01}|}{\sqrt{N_T+\epsilon N_L}} \simeq \frac{|T_{01}|}{\sqrt{|T_{11}|^2+|T_{00}|^2}} \simeq \frac{r^5_{00}}{\sqrt{2r^{04}_{00}}},
\end{equation}
\noindent
where the amplitude $|T_{10}|$
(for the production of transversely polarised $\rho^0$ mesons from 
longitudinal photons) 
is neglected with respect to $|T_{01}|$ and the amplitudes $T_{00}$ 
and $T_{01}$ are 
taken to be in phase; in addition $\epsilon = 1$ is assumed. 
Single and double flip amplitudes 
have been assumed to be small with respect to $T_{00}$ and $T_{11}$.

Finally, the assumption of NPE alone leads to the sum rule:
\begin{eqnarray}\label{NPE}
1-r^{04}_{00} +2 r^{04}_{1-1} -2 r^{1}_{11} - 2 r^{1}_{1-1}=0. 
\end{eqnarray}

\section{Event selection} 

The event selection is similar to that of the previous 
analysis~\cite{rhodis95}; it is summarised briefly in the following.

The online event selection required an 
energy deposit in the BPC or an electron candidate in the CAL,
along with the detection of two tracks in the CTD.

In the offline selection the following further  requirements were imposed:
\begin{itemize}
\item
The energy of the scattered positron was required to be greater than  
20~GeV if  measured  in the BPC and greater than 5 GeV 
if measured   in  the CAL.

\item
The requirement $E-p_Z > 45$~GeV was imposed, where 
$E-p_Z = \sum_i (E_i - p_{Z_i})$ and the summation is over
the energies and longitudinal momenta of the final-state positron
and pions. This cut, applied in the DIS analysis,  
excluded events with high energy photons radiated in the initial state.

\item
The $Z$ coordinate of the interaction vertex was required to
be within $\pm50$ cm of the nominal interaction point.

\item
In addition to the scattered positron, the presence of exactly 
two oppositely charged tracks was required, each associated 
with the  reconstructed vertex, and each with  pseudo\-rap\-i\-di\-ty 
$\left| \eta \right|$ less than 1.75 and  transverse momentum 
greater than 150~MeV; pseudorapidity is defined as
$\eta=-\ln{(\tan{(\theta/2)})}$. These cuts excluded regions of low 
reconstruction efficiency and poor momentum resolution in the CTD. 

\item
Events with any energy deposit larger than 300~MeV in the CAL and not associated 
with either the pion tracks or the positron were rejected~\cite{Heiko,Teresa}.
\end{itemize}

In addition, the following requirements were applied to select 
kinematic  regions of high acceptance:

\begin{itemize}
\item The BPC analysis was limited to the region $0.25<Q^2<0.85$~GeV$^2$ and 
$20 < W < 90$~GeV. 

\item The DIS analysis was restricted to the kinematic region 
$3<Q^2<30$~GeV$^2$ and $40 < W < 120$~GeV.

\item Only events in the $\pi^+\pi^-$ mass  interval  $0.6<M_{\pi\pi}<1.0$~GeV 
and with $|t|<0.6$~GeV$^2$ were taken. 
The mass interval is slightly narrower 
than that used previously~\cite{rhodis95}, in order to reduce the effect of the
background from non-resonant $\pi^+ \pi^-$ production.
In the selected $M_{\pi \pi}$ range the resonant contribution is 
$\approx 98\%$ for the BPC sample and $\approx 100\%$ for the DIS sample.
\end{itemize}

The above selections yielded 5271 events in the BPC sample and
2510 events in the DIS sample.

\section{Monte Carlo simulation}

The relevant Monte Carlo generators have been described in detail 
previously~\cite{rhodis95}. Here their main aspects are summarised.

In the BPC analysis, a dedicated Monte Carlo generator 
based on  the JETSET7.4~\cite{ref:jetset} package
was used. The  simulation of exclusive $\rho^0$ production
was based on the VMD model and Regge phenomenology. 
The effective $Q^2$, $W$ and $t$ dependences of the cross section
were parameterised to reproduce the measurements~\cite{rhodis95}.

In  the DIS analysis, a program~\cite{ref:muchor} interfaced to 
HERACLES4.4~\cite{ref:heracles} was used. Again, the effective $Q^2$, $W$ 
and $t$ dependences of the cross section
were parameterised to reproduce the results~\cite{rhodis95}.

In both cases the decay angular distributions were generated uniformly
and the Monte Carlo events were then iteratively reweighted with 
Eq.~(\ref{full_equation}) using the results 
of the present analysis for the 15 combinations 
of matrix elements $r^{04}_{ik}$, $r^{\alpha}_{ik}$.

\begin{figure}[t] 
\begin{center}
\epsfig{figure=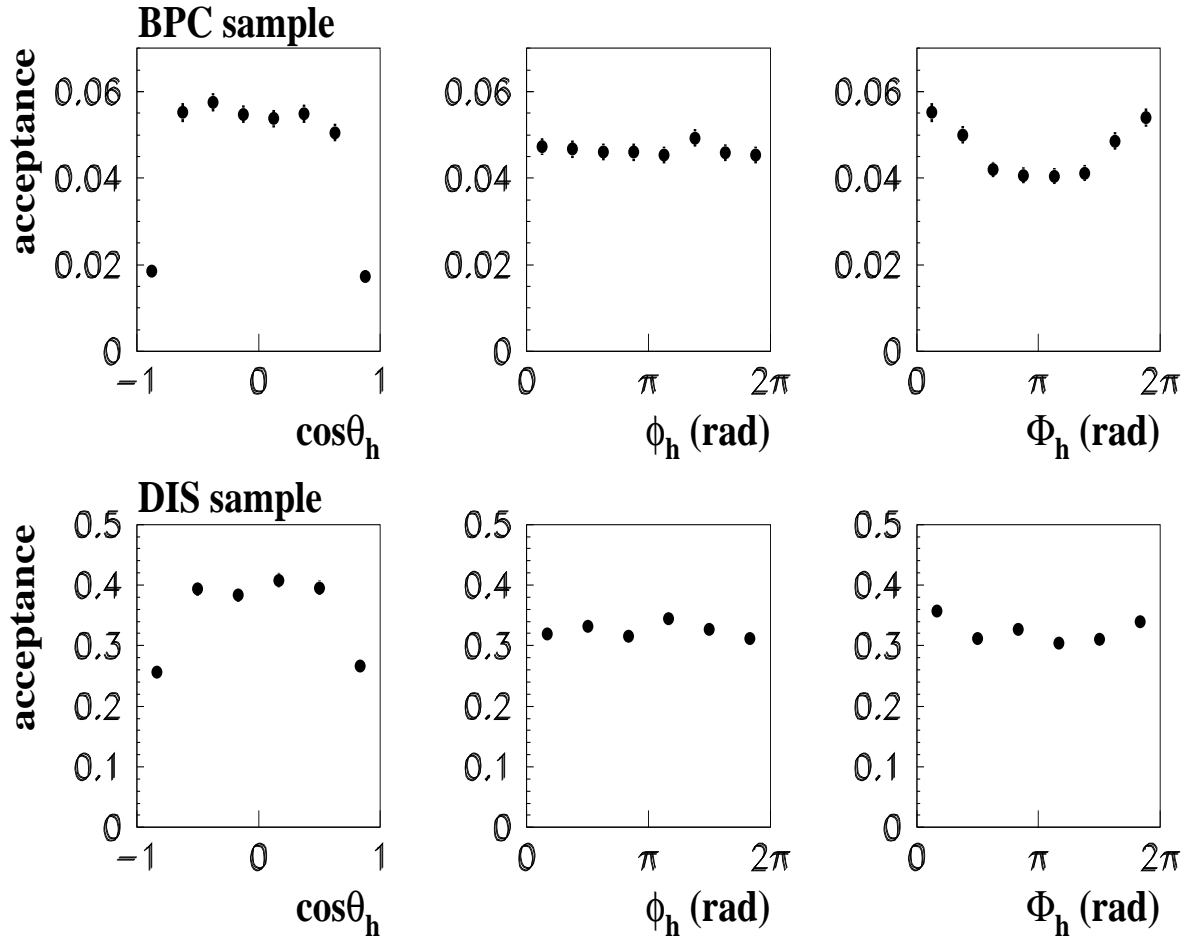,height=15cm,width=17.5cm}
\caption{Acceptance as a function of $\cos{\theta_h}$,
$\phi_h$ and $\Phi_h$, for the BPC and the DIS data samples.
The bars indicate the statistical uncertainties. }
\label{acceptance} 
\end{center} 
\end{figure}

The generated events were processed through the same chain of selection and 
reconstruction procedures as the data, thus accounting for trigger 
as well as detector efficiencies and smearing effects. 
For both analyses, the number of simulated events after reconstruction was  
about a factor of seven greater than the number of reconstructed data events.
Figure~\ref{acceptance} shows the acceptances (defined as in
Ref.~\cite{rhodis95}) as a function of $\cos{\theta_h}$,
$\phi_h$ and $\Phi_h$, both for the BPC and the DIS data. The different 
normalisations of the acceptances for the BPC and DIS samples mainly reflect 
the different coverage for the azimuthal angle of the scattered electron,
the BPC covering only a small fraction of $2\pi$.

\section{Results}

\begin{figure}[t] 
\begin{center}
\epsfig{figure=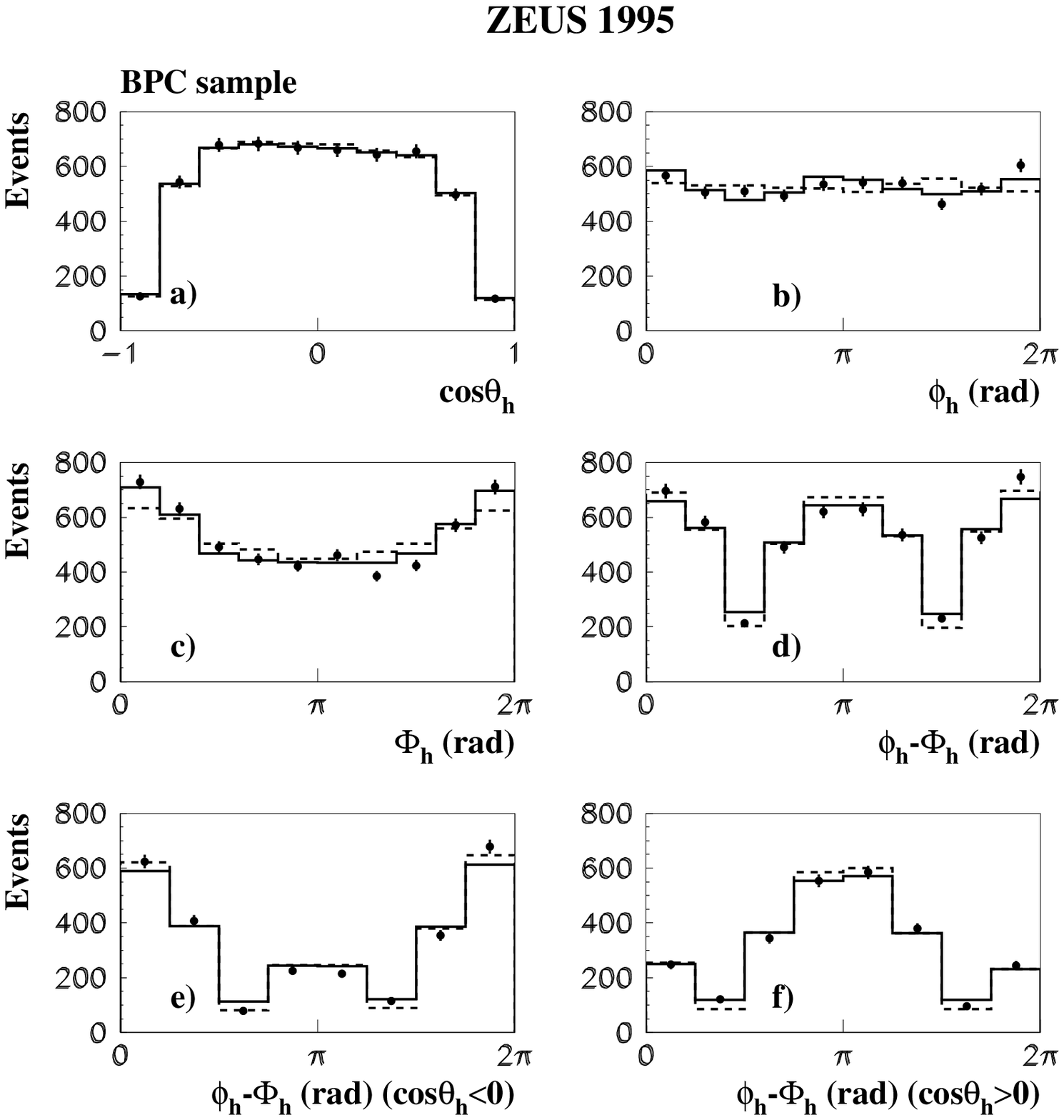,height=18cm}
\caption{Observed distributions for $\cos{\theta_h}$ (a),
$\phi_h$ (b), $\Phi_h$ (c) and $\psi_h=  \phi_h -\Phi_h$ (d, e, f)
of the reconstructed BPC data (points) and the reconstructed 
Monte Carlo events (histograms). The solid histograms 
correspond to the Monte Carlo data reweighted with 
Eq.~(\protect\ref{full_equation}), in which the results 
of the present analysis were used for the 15 combinations 
of matrix elements $r^{04}_{ik}$, $r^{\alpha}_{ik}$. The dashed histograms 
correspond to the SCHC hypothesis. The distributions are not corrected 
for acceptance. The error bars indicate the statistical uncertainty.
The statistical uncertainty of the simulated distributions is negligible.
}
\label{data_mc_bpc} 
\end{center} 
\end{figure}

\begin{figure}[t] 
\begin{center}
\epsfig{figure=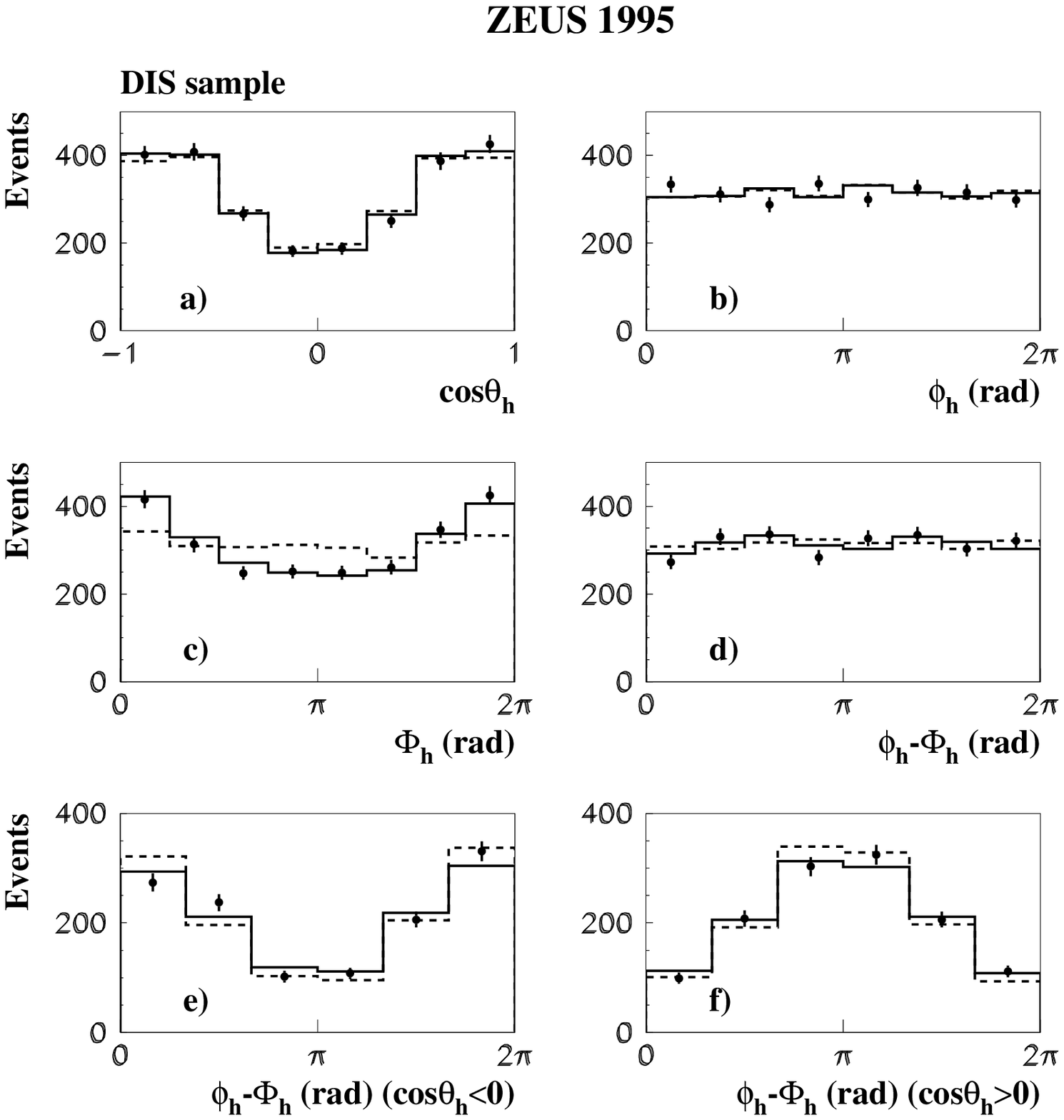,height=18cm}
\caption{Observed distributions for $\cos{\theta_h}$ (a),
$\phi_h$ (b), $\Phi_h$ (c) and $\psi_h=  \phi_h -\Phi_h$ (d, e, f)
of the reconstructed DIS data (points) and the reconstructed 
Monte Carlo events (histograms). The solid histograms 
correspond to the Monte Carlo data reweighted with 
Eq.~(\protect\ref{full_equation}), in which the results 
of the present analysis were used for the 15 combinations 
of matrix elements $r^{04}_{ik}$, $r^{\alpha}_{ik}$. The dashed histograms 
correspond to the SCHC hypothesis. The distributions are not corrected 
for acceptance. The error bars indicate the statistical uncertainty.
The statistical uncertainty of the simulated distributions is negligible.}
\label{data_mc_dis} 
\end{center} 
\end{figure}

Figures~\ref{data_mc_bpc} and~\ref{data_mc_dis} show the reconstructed 
decay angular distributions for the data (full symbols) and Monte Carlo events 
(solid histograms) for the BPC and for the DIS samples, respectively. 
The Monte Carlo simulation reproduces the measured distributions well.
The dashed histograms 
show the same Monte Carlo events reweighted to respect SCHC.
The violation of SCHC can be most clearly seen in the $\Phi_h$ 
distributions, particularly for the DIS sample of Fig.~\ref{data_mc_dis}. 
In the following we focus on how this violation manifests
itself in terms of the 15 combinations of spin-density matrix 
elements, $r^{04}_{ik}$, $r^{\alpha}_{ik}$.

The combinations $r^{04}_{ik}$, $r^{\alpha}_{ik}$ 
were obtained by minimising the difference between the three-dimensional 
($\cos{\theta_h}, \phi_h, \Phi_h$) angular distributions of
the data and those of the simulated events.
A binned likelihood method was used, assuming a Poisson distribution for 
the number of events in each bin. The number of bins in 
$\cos{\theta_h}, \phi_h$ and $\Phi_h$ was $8 \times 8 \times 8$. 

Table~\ref{tab1} summarises the results for the 15
coefficients $r^{04}_{ik}$, $r^{\alpha}_{ik}$. 
The BPC results cover the kinematic region 
$0.25<Q^2<0.85$~GeV$^2$ ($\langle Q^2 \rangle = 0.41$~GeV$^2$), 
$20 < W < 90$~GeV ($\langle W \rangle = 45$~GeV), $0.6<M_{\pi\pi}<1.0$~GeV 
and $|t|<0.6$~GeV$^2$ ($\langle |t| \rangle = 0.14$~GeV$^2$). 
The DIS results cover the region $3<Q^2<30$~GeV$^2$ 
($\langle Q^2 \rangle = 6.3$~GeV$^2$), $40 < W < 120$~GeV 
($\langle W \rangle = 73$~GeV), $0.6<M_{\pi\pi}<1.0$~GeV 
and $|t|<0.6$~GeV$^2$ ($\langle |t| \rangle = 0.17$~GeV$^2$).
Tables~\ref{bpc_w},~\ref{bpc_q} 
and~\ref{bpc_t} give the BPC results in bins of $W$, $Q^2$ and $t$.
The limited statistics and poorer $t$ resolution do not allow 
such fine binning for the DIS sample.
Tables~\ref{tab2} and~\ref{tab3}
give the full correlation matrix for the BPC and the DIS results, 
respectively. 

\begin{figure}[t] 
\begin{center}
\epsfig{figure=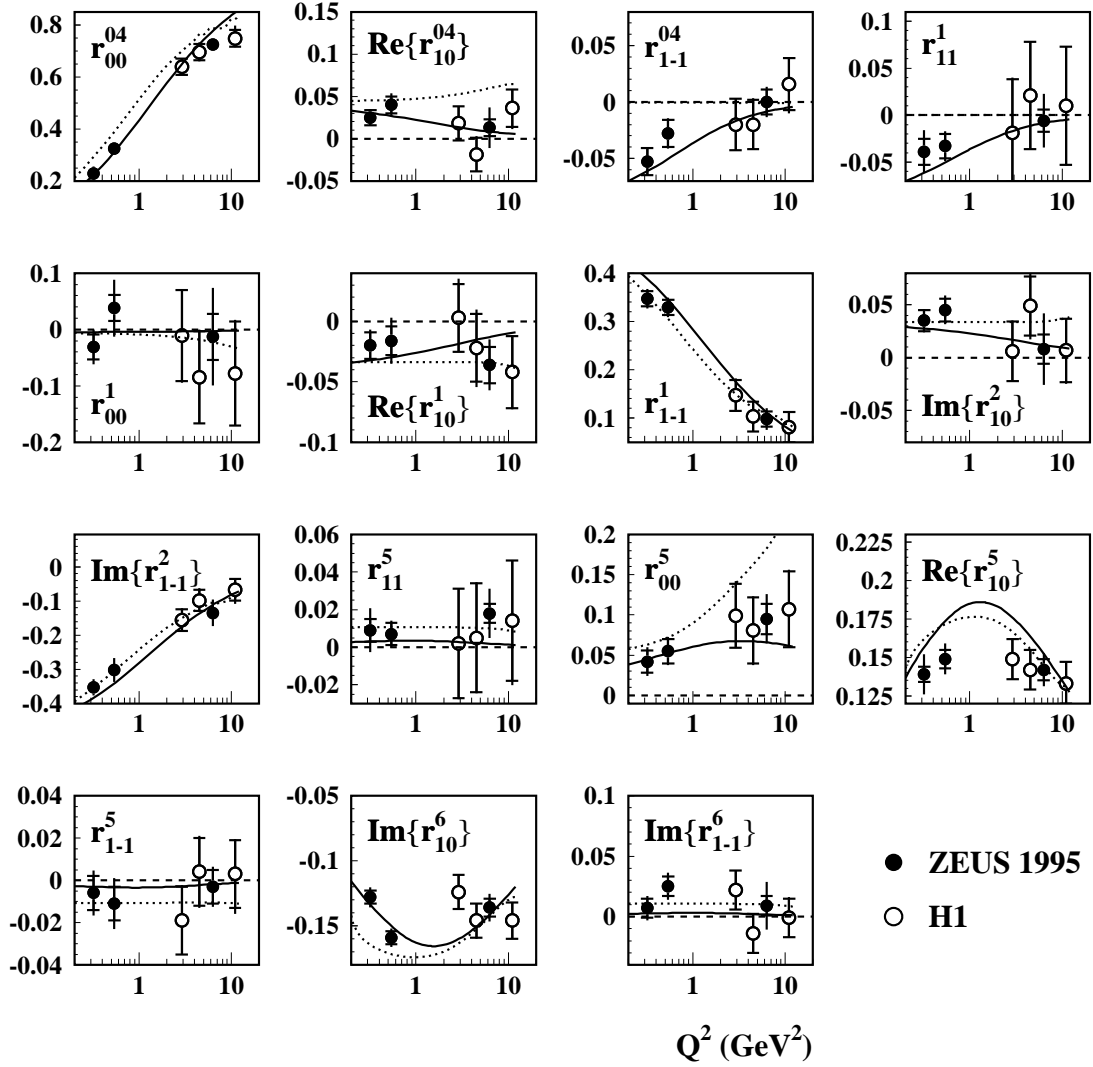,height=18cm}
\caption{The 15 combinations of spin-density matrix elements $r^{04}_{ik}$, 
$r^{\alpha}_{ik}$, 
as obtained from the BPC and DIS data sets as a function of $Q^2$. 
The full symbols indicate the present results; the open points indicate the 
H1 results~\protect\cite{h1}. The inner error bars 
indicate the statistical uncertainties, the outer the statistical and 
systematic uncertainties summed in quadrature. The continuous curves are 
the prediction of the 
model calculation of Ref.~\protect\cite{kolya}, the dotted curves the prediction
of Ref.~\protect\cite{cudell}. The dashed lines indicate the SCHC 
expectation. The predictions of Ref.~\protect\cite{cudell} and of SCHC coincide
for $r^{04}_{1-1}$ and $r^1_{11}$.
}
\label{q2dep} 
\end{center} 
\end{figure}

Figure~\ref{q2dep} shows the combinations of matrix elements for the 
BPC and the DIS data as a function of $Q^2$.
The coefficient $r^{04}_{00}$ represents the probability 
that the $\rho^0$ meson be produced in the helicity zero state, i.e. with 
longitudinal polarisation; it grows with $Q^2$, a result qualitatively in 
agreement with SCHC, since the contribution of longitudinal
photons increases with increasing $Q^2$. Less pronounced variations with 
$Q^2$ are
also observed for $r^1_{1-1}$ and Im$\{r^2_{1-1}\}$, which are related
to $r^{04}_{00}$ under the combined SCHC and NPE assumptions (see 
Sect.~\ref{kinematics}). 

\begin{figure}[t] 
\begin{center}
\epsfig{figure=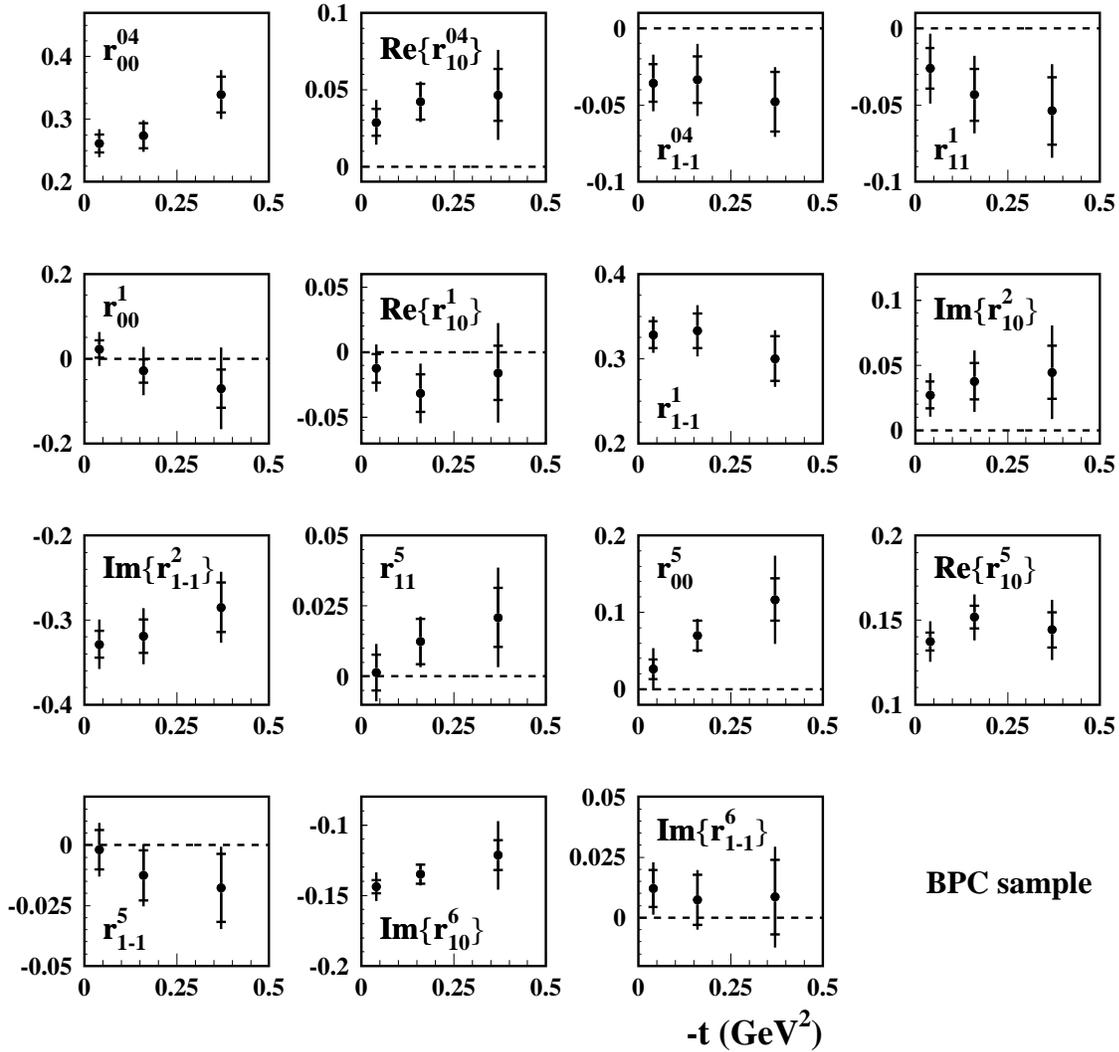,height=18cm}
\caption{The 15 combinations of spin-density matrix elements $r^{04}_{ik}$, 
$r^{\alpha}_{ik}$, 
as obtained from the BPC data as a function of $t$. The inner error bars
indicate the statistical uncertainties, the outer bars the statistical and
systematic uncertainties summed in quadrature. The horizontal bars indicate
the size of the bins. The dashed lines indicate the prediction of SCHC, where 
available. The data cover the kinematic range
$0.25<Q^2<0.85$~GeV$^2$, $20<W<90$~GeV, $0.6<M_{\pi\pi}<1.0$~GeV and 
$|t|<0.6$~GeV$^2$. }
\label{mevst} 
\end{center} 
\end{figure}

Figure~\ref{mevst} shows the 
coefficients $r^{04}_{ik}$, $r^{\alpha}_{ik}$ as 
functions of $t$ for the BPC sample alone. The data suggest a growth of 
$r^5_{00}$ with $|t|$. 

\begin{figure}[t] 
\begin{center}
\epsfig{figure=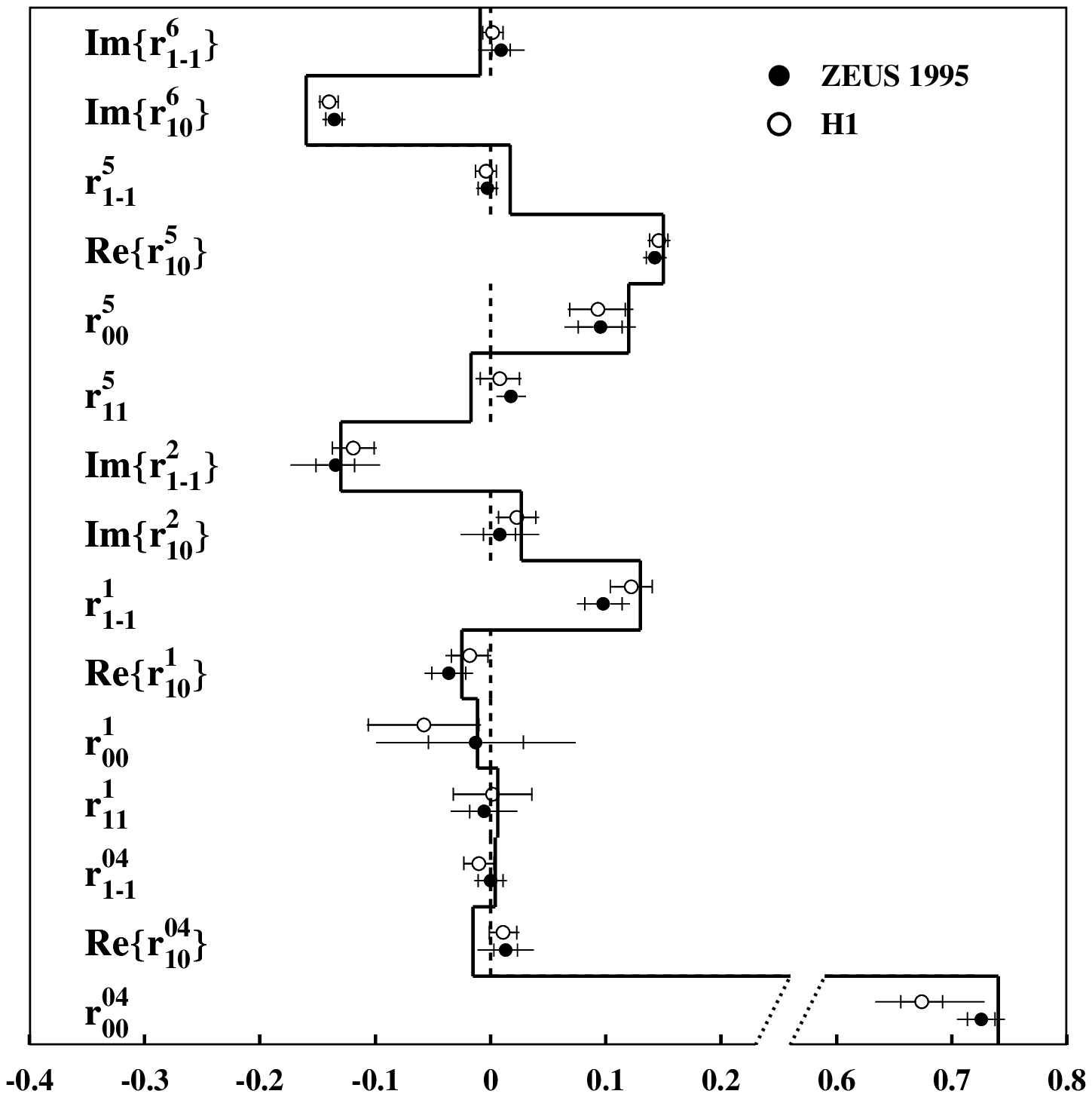,height=17cm}
\caption{The 15 combinations of spin-density matrix elements 
$r^{04}_{ik}$, $r^{\alpha}_{ik}$,
as obtained from the DIS data compared with the predictions
of the model calculation of Ref.~\protect\cite{ivanov} (continuous histogram)
and of SCHC (dashed histogram, see text for details). The solid points
indicate the present results; the open points indicate the H1 data~\protect\cite{h1}.
The inner error bars indicate the statistical 
uncertainties, the outer 
the statistical and systematic uncertainties summed 
in quadrature. The ZEUS data cover the kinematic range
$3<Q^2<30$~GeV$^2$ ($\langle Q^2 \rangle = 6.3$~GeV$^2$), 
$40<W<120$~GeV ($\langle W \rangle = 73$~GeV), $0.6<M_{\pi\pi}<1.0$~GeV and 
$|t|<0.6$~GeV$^2$. The H1 data cover the range 
$1<Q^2<60$~GeV$^2$, $30<W<140$~GeV and $|t|<0.5$~GeV$^2$. 
The prediction of Ref.~\protect\cite{ivanov} is for 
$Q^2=10$~GeV$^2$ and $W=100$~GeV. 
}
\label{ivanov} 
\end{center} 
\end{figure}

Figures~\ref{q2dep},~\ref{mevst} and~\ref{ivanov} compare the results to 
the expectations
of SCHC (dashed lines) for the ten elements for which SCHC makes explicit
predictions. 
Again deviations from SCHC are observed for both the BPC data (Re$\{r^{04}_{10}\}$, 
$r^{04}_{1-1}$, $r^{1}_{11}$, Im$\{r^{2}_{10}\}$, $r^{5}_{00}$) and the DIS 
data ($r^{5}_{00}$). 
A comparison of the results with the SCHC hypothesis yields 
$\chi^2/ndf=81.1/12$ for the BPC sample and
$\chi^2/ndf=79.2/12$ for the DIS sample.
Figures~\ref{q2dep} and~\ref{ivanov} also show the recent H1 
results~\cite{h1}, obtained 
in the kinematic region $1<Q^2<60$~GeV$^2$, $30<W<140$~GeV and
$|t|<0.5$~GeV$^2$; they are in 
excellent agreement with the present data. The results of the model 
calculations~\cite{ivanov,kolya,cudell} are also shown in 
Figs.~\ref{q2dep} and~\ref{ivanov}.

The present results are not corrected for the contribution of non-resonant 
$\pi^+ \pi^-$ production and its
interference with resonant $\rho^0 \rightarrow \pi^+ \pi^-$ production
via the so-called S\"oding mechanism~\cite{soeding}. 
Hence, strictly speaking,
they only apply to the reaction $ep \rightarrow e\pi^+\pi^-p$ with $M_{\pi\pi}$
in the quoted range. While the relative magnitude of the 
non-resonant contribution is small, the interference term, which changes 
sign at the nominal $\rho$ mass value, may significantly affect the 
angular distributions. In order to assess the sensitivity of the data to 
changes in the selected $M_{\pi\pi}$ 
range -- and hence changes in the relative contributions of the
non-resonant and the interference terms -- 
the events with $M_{\pi \pi} < M_{\rho}$ and $M_{\pi \pi}> M_{\rho}$
were analysed separately.
For the DIS sample, where the S\"oding effect is small~\cite{rhodis95,h1}, the 
results thus obtained are consistent with those found for the whole 
$M_{\pi \pi}$ range, suggesting that the observed breaking of SCHC 
is indeed a feature of the reaction $ep \rightarrow e \rho^0 p$.
 The BPC data, however, exhibit some dependence on 
$M_{\pi \pi}$, as shown in Table~\ref{bpcvsmass}. In particular, 
$r^{04}_{00}$ decreases for the higher $M_{\pi \pi}$ selection, a result 
qualitatively 
consistent with expectations based on the interference between resonant 
and non-resonant $\pi^+\pi^-$ production~\cite{ryskin}.
The observed variations of $r^1_{1-1}$ and Im$\{r^2_{1-1}\}$ reflect
that of $r^{04}_{00}$ under the combined SCHC and NPE assumptions (see 
Sect.~\ref{kinematics}). Appreciable $M_{\pi\pi}$ dependences are also observed 
for $r^5_{00}$, Re$\{r^5_{10}\}$ and Im$\{r^6_{10}\}$, which suggests
an influence of the non-resonant $\pi^+ \pi^-$ production on these
combinations of matrix elements.

The results are also not corrected for the contribution 
from the proton-dissociative reaction, $ep \rightarrow e \rho^0 N$, where 
$N$ is a state of mass $M_N$ ($M_N \lsim 4$~GeV) which escapes undetected 
in the proton beam direction. The size of this background 
was estimated~\cite{rhodis95} to be $(23 \pm 8)\%$ and $(24^{+9}_{-5})\%$ 
for the BPC and the DIS samples, respectively. 
The $W$, $Q^2$, $\cos{\theta_h}$, $\phi_h$ and $\Phi_h$ distributions 
for the proton-dissociative events (tagged by activity in the 
FCAL~\cite{rhodis95}) were found to be consistent with those for the exclusive 
events, as expected on the basis of the factorisation of the diffractive 
vertices~\cite{reviews}.
It has therefore been assumed that the proton-dissociative contribution does 
not affect the decay angular distributions. 

Finally, any influence of radiative corrections on the angular distributions
has been neglected.

\section{Systematic errors}

The systematic uncertainties were obtained by modifying the 
analysis procedures as 
listed below.

\begin{enumerate}

\item[(a)] Sensitivity to the event selection criteria:

\begin{itemize}

\item The $M_{\pi\pi}$ range was restricted to $0.7<M_{\pi\pi}<0.84$~GeV
for the BPC sample and $0.65<M_{\pi\pi}<0.9$~GeV for the DIS sample.

\item The minimum positron energy as measured in the BPC was increased to 
23~GeV.

\item The minimum value of $E-P_Z$ was raised to 48~GeV.

\item The minimum track transverse momentum was increased from 150~MeV to 
200~MeV for the BPC analysis and to 300~MeV for the DIS analysis.

\item The pseudorapidity range was restricted to $\left| \eta \right|< 1.7$
for the BPC data and $\left| \eta \right|<1.5$    for the DIS data.

\item The $t$ range was restricted to $0.05<|t|<0.5$~GeV$^2$.

\end{itemize}

\item[(b)] Sensitivity to the Monte Carlo simulation:

\begin{itemize}

\item The $W$ distribution was reweighted by a factor $W^k$,
with $k$ varying between $-0.1$ and 0.1.

\item The $Q^2$ distribution was reweighted by a factor 
$1/(1+Q^2/M_{\rho}^2)^k$, where $k$ varies between $-0.2$ and 0.2 and 
$M_{\rho}$ is the $\rho^0$ meson mass.

\item The $t$ slope was varied by $\pm 1$~GeV$^{-2}$ for the BPC data and
by $\pm 0.5$~GeV$^{-2}$ for the DIS data.
\end{itemize}

\item[(c)] Sensitivity to the fitting procedure: the 
method of moments was used instead of the maximum likelihood method.
Moments of the observed three-dimensional distribution in
the angles $(\cos\theta_{h},\phi_{h}$ and $\Phi_{h})$ 
were calculated, i.e.
the distribution was projected on appropriate functions of 
$\cos{\theta_h}$, $\phi_h$ and $\Phi_h$. The same moments were evaluated 
for the reconstructed angular distribution from the Monte Carlo simulation
and the difference between the moments 
in the data and those in the Monte Carlo was minimised by adjusting
the values of the coefficients $r^{04}_{ik}$, $r^{\alpha}_{ik}$ used 
in the simulation. 
This method is a variation of that presented in 
Appendix C of 
Ref.~\cite{ref:angle}.

\item[(d)] Sensitivity to the binning: the standard number of bins 
in $\cos{\theta_h}$, $\phi_h$ and $\Phi_h$, $8 \times 8 \times 8$,
was
changed to $6 \times 8 \times 8$ bins and, in the BPC case, also to 
$10 \times 10 \times 10$.

\end{enumerate}

The dominant effects come from the sensitivity to the binning, 
the modification of the $M_{\pi\pi}$ range and the 
fitting method (maximum likelihood vs method of moments). As an example, 
for $r^5_{00}$, the systematic uncertainty due to the binning is 
26\% (14\%), and that due to the $M_{\pi\pi}$ range is 15\% (11\%) 
for the BPC (DIS) analysis.

\section{Discussion}
\label{discussion}

In this section the $r^{04}_{ik}$, 
$r^{\alpha}_{ik}$ combinations  which exhibit a deviation from 
SCHC are individually examined. 
The implications of the measured violation of SCHC
for $R=\sigma_{L}^{\gamma^{\star}{p}}/\sigma_{T}^{\gamma^{\star}{p}}$
are also discussed, as is the extent to which NPE dominates the present 
data.

Deviations from the null values expected in the case of SCHC are observed
for Re$\{r^{04}_{10}\}$, $r^{04}_{1-1}$, $r^{1}_{11}$, Im$\{r^{2}_{10}\}$ and
$r^{5}_{00}$:

\begin{enumerate}

\item 
Re$\{r^{04}_{10}\}=0.034 \pm 0.006~\mbox{(stat.)} \pm 0.009~\mbox{(syst.)}$ 
for the BPC sample. The DIS sample gives
Re$\{r^{04}_{10}\}=0.013 \pm 0.010~\mbox{(stat.)} \pm 0.022~\mbox{(syst.)}$,
which is consistent with the BPC result but does not significantly deviate 
from zero. The H1 DIS result, 
Re$\{r^{04}_{10}\}=0.011 \pm 0.012~\mbox{(stat.)}~^{+0.007}_{-0.001}~\mbox{(syst.)}$, 
is consistent with zero~\cite{h1}.

Non-zero values of Re$\{r^{04}_{10}\}$ have also been observed in 
low-energy photoproduction~\cite{ballam}, as well as in 
electroproduction~\cite{joos} and muoproduction~\cite{chio} experiments. 
Reference~\cite{joos} quotes
values ranging from 0.01 to 0.16 for $0 \lsim Q^2 \lsim 1$~GeV$^2$ and 
$1.7<W<2.8$~GeV. At higher energy, $12.5<W<16$~GeV, and for $Q^2<3$~GeV$^2$,
the CHIO Collaboration found Re$\{r^{04}_{10}\}=0.07\pm0.03$~\cite{chio}.

This  coefficient is sensitive (cf. Table~\ref{amplitudes}) 
to the interference of helicity-conserving and 
helicity-single-flip amplitudes, as well as to the interference of 
helicity-single-flip and helicity-double-flip amplitudes. Both 
single-flip amplitudes are involved here, i.e. that for the production of 
longitudinally polarised $\rho^0$ mesons from transverse photons and that
for the production of transversely polarised mesons from longitudinal photons.

The data are consistent with the predictions of~\cite{kolya,cudell} at
low $Q^2$. Some disagreement with the predictions of~\cite{ivanov,cudell}
is visible at high $Q^2$.

\item $r^{04}_{1-1}= - 0.040 \pm 0.009~\mbox{(stat.)}\pm 0.019~\mbox{(syst.)}$ 
for the BPC sample. No deviation from zero is observed for the 
DIS sample, 
$r^{04}_{1-1}= 0.000 \pm 0.011~\mbox{(stat.)}\pm 0.008~\mbox{(syst.)}$.
The H1 DIS result, 
$r^{04}_{1-1}=-0.010 \pm 0.013~\mbox{(stat.)}^{+0.004}_{-0.003}~\mbox{(syst.)}$, 
is consistent with zero~\cite{h1}.

Negative values for $r^{04}_{1-1}$, consistent with the BPC result, though 
only marginally different from zero, were also measured in the range 
$1.7<W<2$~GeV, $Q^2<1$~GeV$^2$~\cite{joos}. 
A value of $r^{04}_{1-1}= - 0.07 \pm 0.03$ was reported
for $12.5<W<16$~GeV~\cite{chio}. Negative values of 
$\rho^0_{1-1}$ were measured in photoproduction~\cite{ballam}.

This term is proportional to the square of the helicity-single-flip
amplitude for the production of transversely polarised $\rho^0$ 
mesons from longitudinal photons, $T_{10}$, 
and is also sensitive to the interference 
of helicity-conserving and helicity-double-flip amplitudes.

For this combination of matrix elements, as for Re$\{r^{04}_{10}\}$ discussed
above, the relative contribution from longitudinal photons is an order
of magnitude larger for the DIS sample than for the BPC
sample, owing to the $Q^2$ dependence of $R$ (see Eq.~\ref{r04}).

The pQCD based models~\cite{ivanov,kolya,cudell} reproduce the DIS data for 
$r^{04}_{1-1}$. At low $Q^2$ some disagreement between 
the model~\cite{cudell} and the data is observed.

\item $r^{1}_{11}=-0.039 \pm 0.009~\mbox{(stat.)} \pm 0.021~\mbox{(syst.)}$ 
for the BPC sample. Here again, no significant effect is visible in the DIS data,
$r^{1}_{11}=-0.006 \pm 0.012~\mbox{(stat.)} \pm 0.026~\mbox{(syst.)}$.
No evidence for a deviation from zero of $r^1_{11}$ was found by the 
fixed-target experiments or by H1; the latter measured 
$r^{1}_{11}=-0.002 \pm 0.034~\mbox{(stat.)} \pm 0.006~\mbox{(syst.)}$
in DIS~\cite{h1}.

This  coefficient is sensitive to the interference between the 
helicity-conserving 
and the helicity-double-flip amplitudes. 

Here also the models~\cite{ivanov,kolya,cudell} reproduce the DIS data.
At low $Q^2$ disagreements similar to those observed for $r^{04}_{1-1}$
between the model~\cite{cudell} and the data are visible.

\item Im$\{r^{2}_{10}\}=0.039 \pm 0.008~\mbox{(stat.)} 
\pm 0.012~\mbox{(syst.)}$ (BPC). The DIS result, 
Im$\{r^{2}_{10}\}=0.008 \pm 0.014~\mbox{(stat.)} \pm 0.031~\mbox{(syst.)}$,
is consistent with zero, 
although also consistent with the BPC result. The H1 DIS result, 
Im$\{r^{2}_{10}\}=0.023 \pm 0.016~\mbox{(stat.)} 
^{+0.010}_{-0.009}~\mbox{(syst.)}$, is also consistent with zero~\cite{h1}.
No significant deviation of 
Im$\{r^{2}_{10}\}$ from zero was found in the earlier experiments.

This coefficient is sensitive to the interference of helicity-conserving and 
helicity-single-flip amplitudes as well as to the interference of 
single-flip and double-flip amplitudes. The single-flip amplitude 
involved is that corresponding to the production of longitudinally polarised 
$\rho^0$ mesons from transverse photons. 

The pQCD based models~\cite{ivanov,kolya,cudell} give a satisfactory description of the
data. 

\item 
$r^5_{00}= 0.051 \pm 0.010~\mbox{(stat.)} \pm 0.018~\mbox{(syst.)} $ (BPC) and 
$r^5_{00}= 0.095 \pm 0.019~\mbox{(stat.)} \pm 0.024~\mbox{(syst.)}$ (DIS). 
The H1 experiment found 
$r^5_{00}= 0.093 \pm 0.024~\mbox{(stat.)} ^{+0.019}_{-0.010}~\mbox{(syst.)}$ 
in DIS~\cite{h1}, in good agreement with the present ZEUS DIS result. 
None of the fixed-target experiments measured significant deviations 
from zero for $r^5_{00}$. 

This coefficient is proportional to the interference between the 
helicity-non-flip amplitude for longitudinal photons $T_{00}$ and the 
helicity-single-flip amplitude $T_{01}$ for the production of longitudinally 
polarised $\rho^0$ mesons from transverse photons.
The BPC data suggest that 
$r^5_{00}$ increases with increasing $|t|$.

The deviation of $r^5_{00}$ from zero is directly related to the 
failure of SCHC to describe the $\Phi_h$ distribution (cf. 
Figs.~\ref{data_mc_bpc} and~\ref{data_mc_dis}), as is apparent by
integrating Eq.~(\ref{full_equation}) over $\cos{\theta_h}$ and $\phi_h$:
\begin{eqnarray}
W(\Phi_h) & \propto & 
[1 -\epsilon\cos{2\Phi_h}(2r^1_{11}+r^1_{00}) + \sqrt{2\epsilon (1+\epsilon)}\cos{\Phi_h} (2r^5_{11}+r^5_{00})]. 
\label{wvsphi}
\end{eqnarray}

The results for $r^5_{00}$ are consistent with the 
predictions of refs.~\cite{ivanov,kolya,cudell}, with the possible exception of
model~\cite{cudell}, which at large $Q^2$ is somewhat higher than the data. 

\end{enumerate}

The SCHC-breaking effects observed are not large. This can be seen, for
instance, from the fact that the SCHC-based relations 
$r_{1-1}^1=-{\rm Im}\{r_{1-1}^2\}$,             
${\rm Re}\{r_{10}^5\}=-{\rm Im}\{r_{10}^6\}$ are 
satisfied within errors:
\begin{eqnarray}\label{schc_ok1}
r^{1}_{1-1} + \mbox{Im}\{r^2_{1-1}\} & = & 0.003 \pm 0.016~\mbox{(stat.)} \pm 0.030~\mbox{(syst.)}~\mbox{(BPC)},\nonumber \\ 
r^{1}_{1-1} + \mbox{Im}\{r^2_{1-1}\} & = & -0.037 \pm 0.028~\mbox{(stat.)}~\pm 0.038~\mbox{(syst.)}~\mbox{(DIS)},\nonumber \\ \nonumber \\
\mbox{Re}\{r^{5}_{10}\} + \mbox{Im}\{r^6_{10}\} & = &  0.001 \pm 0.005~\mbox{(stat.)}\pm 0.013~\mbox{(syst.)}~\mbox{(BPC)},\nonumber \\
\mbox{Re}\{r^{5}_{10}\} + \mbox{Im}\{r^6_{10}\} & = & 0.006 \pm 0.012~\mbox{(stat.)} \pm 0.011~\mbox{(syst.)}~\mbox{(DIS)}.\nonumber   
\end{eqnarray}

The relation $1-r^{04}_{00}-2r^1_{1-1}=0$ (which requires NPE in 
addition to SCHC) is also satisfied, but only within approximately 1.6 
standard deviations (summing statistical and systematic uncertainties in 
quadrature) for the DIS data:

\begin{eqnarray}\label{schc_ok2}
1-r^{04}_{00} -2 r^1_{1-1} & = & 0.060 \pm 0.021~\mbox{(stat.)} \pm 0.047~\mbox{(syst.)}~\mbox{(BPC)},\nonumber \\
1-r^{04}_{00} -2 r^1_{1-1} & = & 0.079 \pm 0.032~\mbox{(stat.)} \pm 0.036~\mbox{(syst.)}~\mbox{(DIS)}. \nonumber 
\end{eqnarray}

\noindent

The ratio of the $\rho^0$ electroproduction cross section
for longitudinal photons to that for transverse photons, $R$, was also 
determined under a much weaker set of assumptions than SCHC 
(Eqs.~(\ref{eq:R_ratio_no_schc})-(\ref{eq:delta})). The BPC data give 
$R=0.368 ^{+0.021}_{-0.020}~\mbox{(stat.)}~^{+0.024}_{-0.023}~\mbox{(syst.)}$, 
to be compared with the SCHC-based value
$R=0.377 \pm 0.021~\mbox{(stat.)}~\pm 0.021~\mbox{(syst.)}$.
The DIS data yield 
$R=2.58^{+0.16}_{-0.15}~\mbox{(stat.)}~^{+0.24}_{-0.21}~\mbox{(syst.)}$
while the SCHC-based value is
$R=2.66 ^{+0.17}_{-0.15}~\mbox{(stat.)}~^{+0.24}_{-0.21}\mbox{(syst.)}$.
The new results differ from
the value derived from the SCHC hypothesis by less than 3\% when averaged
over either the BPC or DIS data samples. It can therefore be concluded that 
the results for
$R$ presented in Ref.\cite{rhodis95} are not significantly influenced by the
SCHC assumption used in obtaining them.

The size of the SCHC-breaking effects can also be quantified by evaluating
the ratios of the helicity-single-flip and helicity-double-flip amplitudes 
to the helicity-conserving amplitudes:

\begin{itemize}

\item The ratio of $T_{01}$ (for the production of longitudinally polarised 
$\rho^0$ mesons from transverse photons) to the helicity-conserving amplitudes
can be estimated as (cf. Eq.~(\ref{eq:delta}))
\begin{eqnarray}\label{single_flip}
\tau_{01} = \frac{|T_{01}|}{\sqrt{|T_{00}|^2 + |T_{11}|^2}} \simeq \frac{r^5_{00}}{\sqrt{2 r^{04}_{00}}},
\end{eqnarray}
\noindent
which holds under the same assumptions specified for Eq.~(\ref{eq:delta}).
Equation~(\ref{single_flip}) gives $\tau_{01}=(6.9 \pm 1.3~\mbox{(stat.)}
\pm 2.4~\mbox{(syst.)})\%$ for the BPC data and 
$\tau_{01}=(7.9~\pm 1.6~\mbox{(stat.)}~\pm 2.0~\mbox{(syst.)})\%$ for the 
DIS data. The H1 result for this quantity is $(8\pm 3)\%$~\cite{h1}.

\item Likewise, the ratio
\begin{eqnarray}\label{single_flip_LT}
\tau_{10}= \frac{|T_{10}|}{\sqrt{|T_{00}|^2 + |T_{11}|^2}} 
\simeq \frac{|\mbox{Re}\{r^{04}_{10}\} + \mbox{Re}\{r^{1}_{10}\}|}{\sqrt{r^{04}_{00}}}
\end{eqnarray}
\noindent
can be evaluated, where $T_{10}$ is the amplitude for the 
production of transversely polarised $\rho^0$ mesons from longitudinal photons.
Equation~(\ref{single_flip_LT}) is exact if NPE holds and $T_{00}$ and 
$T_{10}$ have the same phase.
The value $\tau_{10}=(2.9 \pm 1.4~\mbox{(stat.)}~\pm 3.0~\mbox{(syst.)})\%$
is found
for the BPC sample and $\tau_{10}=(2.7 \pm 1.9~\mbox{(stat.)}~\pm 3.1~\mbox{(syst.)})\%$ 
for the DIS data. These results are 
consistent with the expectation of
a severe suppression of the production of transversely polarised $\rho^0$ 
mesons from longitudinal photons~\cite{diehl}. 
 
\item The size of the helicity-double-flip amplitude $T_{1-1}$ can be 
derived assuming NPE and the same phase for $T_{1-1}$  and $T_{11}$:
\begin{eqnarray}\label{double_flip}
\tau_{1-1}= \frac{|T_{1-1}|}{\sqrt{|T_{00}|^2 + |T_{11}|^2}} 
\simeq \frac{|r^{1}_{11}|}{\sqrt{2r^{1}_{1-1}}},
\end{eqnarray}
\noindent
which gives $\tau_{1-1}=(4.8 \pm 1.1~\mbox{(stat.)}~\pm 2.6~\mbox{(syst.)})\%$
in the BPC analysis and 
$\tau_{1-1}=(1.4 \pm 2.7~\mbox{(stat.)}~\pm 5.9~\mbox{(syst.)})\%$
for the DIS sample.

\end{itemize}

Finally, the data can be used to verify the validity of the NPE hypothesis.
The left-hand-side of Eq.~(\ref{NPE}) is
$0.058 \pm 0.027~\mbox{(stat.)} \pm 0.074~\mbox{(syst.)}$ for the 
BPC sample and $0.091 \pm 0.043~\mbox{(stat.)} \pm 0.065~\mbox{(syst.)}$ 
for the DIS sample. Both results are consistent with the NPE expectation of zero.

\section{Summary and conclusions}

Exclusive electroproduction of $\rho^0$ mesons has been measured at HERA 
in two $Q^2$ ranges, $0.25<Q^2<0.85$~GeV$^2$ and 
$3<Q^2<30$~GeV$^2$, and
the angular distributions of the decay pions have been studied.
 The low-$Q^2$ data span the range 
$20<W<90$~GeV; the high-$Q^2$ data cover the
$40<W<120$~GeV interval. Both samples extend up to $|t|=0.6$~GeV$^2$. 
The available statistics enable a determination of the 
15 combinations of spin-density matrix elements, $r^{04}_{ik}$ and
$r^{\alpha}_{ik}$, and allow a check of the extent to which 
$s$-channel helicity conservation holds at HERA energies for this process.
An overall comparison of the data 
with the SCHC hypothesis leads to a very poor
$\chi^2$, $\chi^2/ndf=81.1/12$, for the BPC sample and  
$\chi^2/ndf=79.2/12$ for the DIS sample. The deviation from SCHC is most evident
in the $\Phi_h$ distribution.

In the low-$Q^2$ sample, five of the $r^{04}_{ik}$ and $r^{\alpha}_{ik}$ 
combinations deviate from the null value expected if SCHC is assumed. 
The combination $r^5_{00}$ is measured to be
$r^5_{00}= 0.051 \pm 0.010~\mbox{(stat.)} \pm 0.018~\mbox{(syst.)}$,
which indicates the production of longitudinally
polarised (i.e. helicity zero) $\rho^0$ mesons from transverse photons.
In addition,
Re$\{r^{04}_{10}\}=0.034 \pm 0.006~\mbox{(stat.)} \pm 0.009~\mbox{(syst.)}$,
$r^{04}_{1-1}= - 0.040 \pm 0.009~\mbox{(stat.)}\pm 0.019~\mbox{(syst.)}$,
$r^{1}_{11}=-0.039 \pm 0.009~\mbox{(stat.)} \pm 0.021~\mbox{(syst.)}$ and
Im$\{r^{2}_{10}\}=0.039 \pm 0.008~\mbox{(stat.)} 
\pm 0.012~\mbox{(syst.)}$.
The measured values for Re$\{r^{04}_{10}\}$ and $r^{04}_{1-1}$ are consistent
with the results of earlier fixed-target experiments. Some variation of these 
results with $M_{\pi\pi}$ is observed, possibly indicating a contribution
from non-resonant $\pi^+ \pi^-$ production and its interference with
resonant $\rho^0$ production.

In the high-$Q^2$ sample, the only coefficient significantly different from the 
SCHC expectation is $r^5_{00}$, for which the value
$r^5_{00}= 0.095 \pm 0.019~\mbox{(stat.)} \pm 0.024~\mbox{(syst.)}$ is measured.
This value of $r^5_{00}$ is consistent with the predictions of the pQCD based 
models~\cite{ivanov,kolya,cudell}. The results for all coefficients
$r^{04}_{ik}$ and $r^{\alpha}_{ik}$ at large $Q^2$ are in good agreement with 
those found by H1~\cite{h1}.

The ratio $R$ of the cross sections for $\rho^0$ production 
from longitudinal and transverse photons was also determined without assuming 
SCHC. The results thus found differ from 
those derived from the SCHC hypothesis by less than 3\% when averaged
over either the BPC or DIS data samples. The results for
$R$ presented in Ref.\cite{rhodis95} are thus not significantly influenced by 
the SCHC assumption used in obtaining them.

Finally, the data support the hypothesis of natural parity exchange in 
$\rho^0$ electroproduction.

In conclusion, a small breaking of SCHC, less than 10\% in the
amplitudes, is a characteristic feature 
of exclusive $\rho^0$ meson production also at the large values of $W$ accessible 
at HERA. In the high-$Q^2$ region, the effect is quantitatively reproduced by 
perturbative QCD calculations.

\section*{Acknowledgements}
We thank the DESY Directorate for their strong support and encouragement,
and the HERA machine group for their diligent efforts.
We are grateful for the  support of the DESY computing
and network services. The design, construction and installation
of the ZEUS detector have been made possible by the ingenuity and effort
of many people from DESY and home institutes who are not listed
as authors. It is also a pleasure to thank I.~Akushevich, M. Diehl, 
D.Yu.~Ivanov, 
N.N.~Nikolaev, I.~Royen, M.G.~Ryskin, Yu.M. Shabelski and B.G.~Zakharov for many 
useful discussions. We are grateful to I.~Akushevich and I. Royen for 
calculating the predictions of their models for the kinematic range covered
by our data.

\clearpage
\begin{table}
\begin{tabular}{l}
$r^{04}_{00}=\frac{\eps |T_{00}|^2}{\DR}+\frac{|T_{01}|^2}{\DR}$\\
\\
$\R\{r^{04}_{10}\}=\frac{1}{2}\frac{\R\{T_{11} T^{\star}_{01}\}}{\DR}
+\frac{\eps\R\{T_{10}T^{\star}_{00}\}}{\DR}
+\frac{1}{2}\frac{\R\{T_{1-1}T^{\star}_{0-1}\}}{\DR}$\\
\\
$r^{04}_{1-1}=-\frac{\eps |T_{10}|^2}{\DR}+\frac{\R\{T_{11}T^{\star}_{1-1}\}}{\DR}$\\
\\
$r^{1}_{11} = \frac{\R\{T_{1-1}T^{\star}_{11}\}}{\DR}$\\
\\
$r^{1}_{00} =-\frac{ |T_{01}|^2}{\DR}$\\
\\
$\R\{r^{1}_{10}\}=\frac{1}{2} \frac{\R\{T_{11}T^{\star}_{0-1}\}}{\DR}
+\frac{1}{2} \frac{\R\{T_{1-1}T^{\star}_{01}\}}{\DR}$\\
\\
$r^{1}_{1-1} = \frac{1}{2} \frac{ |T_{11}|^2}{\DR}
+\frac{1}{2} \frac{ |T_{1-1}|^2}{\DR}$\\
\\
$\I\{r^{2}_{10}\}=-\frac{1}{2}\frac{\R\{T_{11}T^{\star}_{0-1}\}}{\DR}
+\frac{1}{2}\frac{\R\{T_{1-1}T^{\star}_{01}\}}{\DR}$\\
\\
$\I\{r^{2}_{1-1}\}=-\frac{1}{2} \frac{ |T_{11}|^2}{\DR}
+\frac{1}{2} \frac{ |T_{1-1}|^2}{\DR}$\\
\\
$r^{5}_{11}=\frac{1}{\sqrt{2}}\frac{\R\{T_{10}T^{\star}_{11}\}}{\DR}
-\frac{1}{\sqrt{2}}\frac{\R\{T_{10}T^{\star}_{1-1}\}}{\DR}$\\
\\
$r^{5}_{00}=\frac{\sqrt{2} \R\{T_{00}T^{\star}_{01}\}}{\DR}$\\
\\
$\R\{r^{5}_{10}\}=\frac{1}{2\sqrt{2}}\frac{\R\{T_{11}T^{\star}_{00}\}}{\DR}
+\frac{1}{\sqrt{2}}\frac{\R\{T_{10}T^{\star}_{01}\}}{\DR}
-\frac{1}{2\sqrt{2}}\frac{\R\{T_{1-1}T^{\star}_{00}\}}{\DR}$\\
\\
$r^{5}_{1-1}=\frac{1}{\sqrt{2}}\frac{\R\{T_{11}T^{\star}_{-10}\}}{\DR}
+\frac{1}{\sqrt{2}}\frac{\R\{T_{10}T^{\star}_{-11}\}}{\DR}$\\
\\
$\I\{r^{6}_{10}\}=-\frac{1}{2\sqrt{2}}\frac{\R\{T_{11}T^{\star}_{00}\}}{\DR}
-\frac{1}{2\sqrt{2}}\frac{\R\{T_{1-1}T^{\star}_{00}\}}{\DR}$\\
\\
$\I\{r^{6}_{1-1}\}=-\frac{1}{\sqrt{2}}\frac{\R\{T_{-10}T^{\star}_{11}\}}{\DR}
+\frac{1}{\sqrt{2}}\frac{\R\{T_{10}T^{\star}_{-11}\}}{\DR}$\\
\\
where 
$N_L=|T_{00}|^2+2|T_{10}|^2~~~~~~~N_T=|T_{11}|^2+|T_{1-1}|^2+|T_{01}|^2$\\
\end{tabular}
\vspace{1cm}
\caption{The 15 combinations of spin-density matrix elements 
$r^{04}_{ik}$, $r^{\alpha}_{ik}$, expressed in terms of the 
helicity amplitudes; natural parity exchange in 
the $t$ channel is assumed for $r^{04}_{1-1}$, $r^1_{00}$, $r^1_{1-1}$,
Im$\{r^2_{1-1}\}$, Re$\{r^5_{10}\}$, Im$\{r^6_{10}\}$.
}
\label{amplitudes}
\end{table}

%
%
\clearpage
\begin{table}
\renewcommand{\arraystretch}{1.3}
\def\largestrut{\vrule width 0pt height 30pt depth 20pt\relax}
\begin{center}
\begin{tabular}{|c|r@{$\ \pm\ $}l@{$\ \pm\ $}l|r@{$\ \pm\ $}l@{$\ \pm\ $}l|}
\hline
& \multicolumn{3}{c|}{ZEUS 1995 (BPC) } 
& \multicolumn{3}{c|}{ZEUS 1995 (DIS) }   \\ 
\hline
$r^{04}_{00}$&     $0.272$&$0.011$&$0.011$   &$0.725$&$0.012$ &$0.017$   \\
\hline
Re\{$r^{04}_{10}$\}& $0.034$&$0.006$&$0.009$   &$0.013$&$0.010$ &$0.022$  \\
\hline
$r^{04}_{1-1}  $&  $-0.040$&$0.009$&$0.019$  &$0.000$&$0.011$ &$0.008$  \\
\hline
$r^1_{11}    $&    $-0.039$&$0.009$& $0.021$ &$-0.006$&$0.012$ &$0.026$ \\
\hline
$r^1_{00}    $&    $0.004$&$0.015$& $0.038$  &$-0.013$&$0.041$ &$0.076$ \\
\hline
Re\{$r^1_{10}$\}&    $-0.019$&$0.008$&$0.013$  &$-0.036$&$0.015$ &$0.015$ \\ 
\hline
$r^1_{1-1}   $&    $0.334$&$0.011$& $0.023$  &$0.098$&$0.016$ &$0.016$   \\
\hline
Im\{$r^2_{10}  $\}&  $0.039$&$0.008$& $0.012$  &$0.008$&$0.014$ &$0.031$  \\
\hline
Im\{$r^2_{1-1} $\}&  $-0.331$&$0.011$& $0.020$ &$-0.135$&$0.017$ &$0.035$ \\
\hline
$r^5_{11}    $&    $0.008$&$0.004$& $0.012$  &$0.018$&$0.005$ &$0.012$ \\
\hline
$r^5_{00}    $&    $0.051$&$0.010$& $0.018$  & $0.095$&$0.019$ &$0.024$ \\
\hline
Re\{$r^5_{10}$\}  &  $0.142$&$0.004$& $0.012$  &$0.142$&$0.007$ &$0.008$ \\
\hline
$r^5_{1-1}   $&    $-0.010$&$0.006$& $0.008$ &$-0.003$&$0.008$ &$0.005$  \\
\hline
Im\{$r^6_{10}  $\}&  $-0.141$&$0.003$& $0.005$ &$-0.136$&$0.007$ &$0.008$  \\
\hline
Im\{$r^6_{1-1} $\}&  $0.014$&$0.006$& $0.007$  &$0.009$&$0.008$ &$0.018$  \\
\hline 
\end{tabular}
\vspace{1cm}
\caption{
The 15 combinations of spin-density matrix elements $r^{04}_{ik}$, 
$r^{\alpha}_{ik}$,
as obtained from the BPC and DIS data sets. Statistical
and systematic uncertainties are given separately. 
The BPC data cover the kinematic range
$0.25<Q^2<0.85$~GeV$^2$ ($\langle Q^2 \rangle = 0.41$~GeV$^2$), 
$20<W<90$~GeV ($\langle W \rangle = 45$~GeV), $0.6<M_{\pi\pi}<1.0$~GeV and 
$|t|<0.6$~GeV$^2$ ($\langle |t| \rangle = 0.14$~GeV$^2$). 
The DIS data cover the kinematic range
$3<Q^2<30$~GeV$^2$ ($\langle Q^2 \rangle = 6.3$~GeV$^2$), $40<W<120$~GeV
($\langle W \rangle = 73$~GeV), $0.6<M_{\pi\pi}<1.0$~GeV and 
$|t|<0.6$~GeV$^2$ ($\langle |t| \rangle = 0.17$~GeV$^2$).
}
\label{tab1}
\end{center}
\end{table}

\clearpage
%
%
\clearpage
\begin{table}
\renewcommand{\arraystretch}{1.3}
\def\largestrut{\vrule width 0pt height 30pt depth 20pt\relax}
\begin{center}
\begin{tabular}{|c|r@{$\ \pm\ $}l@{$\ \pm\ $}l|r@{$\ \pm\ $}l@{$\ \pm\ $}l|}
\hline
 \multicolumn{7}{|c|}{ZEUS 1995 (BPC) }  \\ 
\hline

& \multicolumn{3}{c|}{$20<W<45$ GeV } 
& \multicolumn{3}{c|}{$45<W<90$ GeV }   \\ 
\hline
$r^{04}_{00}$&  $ 0.293$&$ 0.014$&$ 0.016$    &  $ 0.257$&$ 0.017$&$ 0.021 $ \\
\hline
Re\{$r^{04}_{10}$\}& $0.034$&$ 0.009$&$ 0.008 $ &  $0.027$&$ 0.010$&$ 0.005 $ \\
\hline
$r^{04}_{1-1}  $&  $-0.032$&$ 0.011$&$ 0.009 $ & $-0.056$&$ 0.013$&$ 0.020 $  \\
\hline
$r^1_{11}    $&   $ -0.040$&$ 0.012$&$ 0.006$ &  $-0.038$&$ 0.015$&$ 0.007 $ \\
\hline
$r^1_{00}    $&   $0.004$&$ 0.021$&$ 0.030 $ &   $0.003$&$ 0.025$&$ 0.051 $ \\
\hline
Re\{$r^1_{10}$\}&   $ -0.024$&$ 0.011$&$ 0.016$ &  $ -0.012$&$ 0.013$&$ 0.005$ \\ 
\hline
$r^1_{1-1}   $&   $0.328$&$ 0.015$&$ 0.014 $  &  $0.345$&$ 0.018$&$ 0.018 $   \\
\hline
Im\{$r^2_{10}  $\}& $0.037$&$ 0.010$&$ 0.013 $ &   $0.038$&$ 0.012$&$ 0.006 $  \\
\hline
Im\{$r^2_{1-1} $\}& $-0.313$&$ 0.015$&$ 0.023 $ &  $-0.348$&$ 0.018$&$ 0.022 $ \\
\hline
$r^5_{11}    $&   $0.015$&$ 0.006$&$ 0.008 $  &  $ 0.004$&$ 0.007$&$ 0.010$ \\
\hline
$r^5_{00}    $&   $0.039$&$ 0.013$&$ 0.016 $  &  $ 0.050$&$ 0.016$&$ 0.007$ \\
\hline
Re\{$r^5_{10}$\}  & $0.143$&$ 0.005$&$ 0.008 $   & $0.142$&$ 0.006$&$ 0.013 $ \\
\hline
$r^5_{1-1}   $&   $-0.013$&$ 0.008$&$ 0.007 $  & $ -0.008$&$ 0.009$&$ 0.010$  \\
\hline
Im\{$r^6_{10}  $\}& $-0.150$&$ 0.004$&$ 0.004 $  & $-0.131$&$ 0.006$&$ 0.007 $  \\
\hline
Im\{$r^6_{1-1} $\}& $ 0.011$&$ 0.007$&$ 0.004 $  & $ 0.022$&$ 0.009$&$ 0.007$  \\
\hline 
\end{tabular}
\vspace{1cm}
\caption{
The 15 combinations of spin-density matrix elements $r^{04}_{ik}$, 
$r^{\alpha}_{ik}$,
as obtained from the BPC data in two $W$ intervals. Statistical
and systematic uncertainties are given separately. 
The data cover the kinematic range
$0.25<Q^2<0.85$~GeV$^2$, $20<W<90$~GeV, $0.6<M_{\pi\pi}<1.0$~GeV and 
$|t|<0.6$~GeV$^2$. The average $W$ values for the two intervals are 
$\langle W \rangle=31$~GeV and $\langle W \rangle=61$~GeV, respectively.
}
\label{bpc_w}
\end{center}
\end{table}
%
%

%
%
\clearpage
\begin{table}
\renewcommand{\arraystretch}{1.3}
\def\largestrut{\vrule width 0pt height 30pt depth 20pt\relax}
\begin{center}
\begin{tabular}{|c|r@{$\ \pm\ $}l@{$\ \pm\ $}l|r@{$\ \pm\ $}l@{$\ \pm\ $}l|}
\hline
\multicolumn{7}{|c|}{ZEUS 1995 (BPC) }  \\
\hline
& \multicolumn{3}{c|}{$0.25<Q^2<0.4$ GeV$^2$ } 
& \multicolumn{3}{c|}{$0.4<Q^2<0.85$ GeV$^2$ }   \\ 
\hline
$r^{04}_{00}$&     $0.228$&$ 0.016$&$ 0.022 $ &  $0.324$&$ 0.015$&$ 0.012 $ \\
\hline
Re\{$r^{04}_{10}$\}& $0.025$&$ 0.009$&$ 0.004 $ &  $0.040$&$ 0.010$&$ 0.009 $ \\
\hline
$r^{04}_{1-1}  $&  $-0.053$&$ 0.012$&$ 0.003 $ & $-0.028$&$ 0.012$&$ 0.006 $ \\
\hline
$r^1_{11}    $&    $-0.039$&$ 0.014$&$ 0.018 $ & $-0.033$&$ 0.013$&$ 0.009 $ \\
\hline
$r^1_{00}    $&    $-0.031$&$ 0.022$&$ 0.022 $ & $0.038$&$ 0.023$&$ 0.045 $ \\
\hline
Re\{$r^1_{10}$\}&    $-0.020$&$ 0.011$&$ 0.007 $ & $-0.016$&$ 0.012$&$ 0.015 $\\ 
\hline
$r^1_{1-1}   $&    $0.347$&$ 0.016$&$ 0.013 $  & $0.329$&$ 0.016$&$ 0.012 $  \\
\hline
Im\{$r^2_{10}  $\}&  $0.035$&$ 0.010$&$ 0.005 $ &  $0.045$&$ 0.011$&$ 0.008 $  \\
\hline
Im\{$r^2_{1-1} $\}&  $-0.353$&$ 0.016$&$ 0.018 $ & $-0.302$&$ 0.016$&$ 0.031 $ \\
\hline
$r^5_{11}    $&    $0.009$&$ 0.006$&$ 0.010 $  & $0.007$&$ 0.006$&$ 0.004 $ \\
\hline
$r^5_{00}    $&    $0.042$&$ 0.014$&$ 0.011 $  & $0.055$&$ 0.015$&$ 0.009 $ \\
\hline
Re\{$r^5_{10}$\}  &  $0.139$&$ 0.005$&$ 0.012 $   &$0.149$&$ 0.006$&$ 0.008 $ \\
\hline
$r^5_{1-1}   $&    $-0.006$&$ 0.008$&$ 0.007 $  &$-0.011$&$ 0.008$&$ 0.009 $ \\
\hline
Im\{$r^6_{10}  $\}&  $-0.128$&$ 0.005$&$ 0.006 $  &$-0.159$&$ 0.005$&$ 0.005 $ \\
\hline
Im\{$r^6_{1-1} $\}&  $0.007$&$ 0.008$&$ 0.006  $  &$0.025$&$ 0.008$&$ 0.008 $  \\
\hline 
\end{tabular}
\vspace{1cm}
\caption{
The 15 combinations of spin-density matrix elements $r^{04}_{ik}$, $r^{\alpha}_{ik}$, 
as obtained from the BPC data in two $Q^2$ intervals. Statistical
and systematic uncertainties are given separately. 
The data cover the kinematic range
$0.25<Q^2<0.85$~GeV$^2$, $20<W<90$~GeV, $0.6<M_{\pi\pi}<1.0$~GeV and 
$|t|<0.6$~GeV$^2$. The average $Q^2$ values for the two intervals are 
$\langle Q^2 \rangle=0.32$~GeV$^2$ and $\langle Q^2 \rangle=0.54$~GeV$^2$, 
respectively.
}
\label{bpc_q}
\end{center}
\end{table}
\clearpage
%
%
\clearpage
\begin{table}
\renewcommand{\arraystretch}{1.3}
\def\largestrut{\vrule width 0pt height 30pt depth 20pt\relax}
\begin{center}
\begin{tabular}{|c|r@{$\ \pm\ $}l@{$\ \pm\ $}l|r@{$\ \pm\ $}l@{$\ \pm\ $}l|r@{$\ \pm\ $}l@{$\ \pm\ $}l|}
\hline
\multicolumn{10}{|c|}{ZEUS 1995 (BPC) }  \\
\hline
& \multicolumn{3}{c|}{$0<|t|<0.1$ GeV$^2$} 
& \multicolumn{3}{c|}{$0.1<|t|<0.25$ GeV$^2$} 
& \multicolumn{3}{c|}{$0.25<|t|<0.6$ GeV$^2$} \\ 
\hline
$r^{04}_{00}$&     $0.261$&$0.014$&$0.017$& $0.273$&$0.020$&$0.016$&  $0.339$&$ 0.029$&$ 0.027 $   \\
\hline
Re\{$r^{04}_{10}$\}& $0.029$&$0.009$&$0.012$& $0.042$&$0.012$&$0.006$&  $0.047$&$ 0.017$&$ 0.024 $ \\
\hline
$r^{04}_{1-1}  $& $-0.036$&$0.012$&$0.014$& $-0.034$&$0.015$&$0.018$& $-0.048$&$ 0.019$&$ 0.012 $  \\
\hline
$r^1_{11}    $&   $-0.026$&$0.013$&$0.018$& $-0.043$&$0.017$&$0.019$& $-0.054$&$ 0.022$&$ 0.021 $ \\
\hline
$r^1_{00}    $&    $0.023$&$0.020$&$0.034$& $-0.029$&$0.028$&$0.051$& $-0.071$&$ 0.045$&$ 0.086 $ \\
\hline
Re\{$r^1_{10}$\}&   $-0.012$&$0.011$&$0.014$& $-0.032$&$0.014$&$0.018$& $-0.016$&$ 0.021$&$ 0.032 $ \\
\hline
$r^1_{1-1}   $&   $0.328$&$0.016$&$0.015$&  $0.333$&$0.020$&$0.023$&  $0.300$&$ 0.026$&$ 0.020 $    \\
\hline
Im\{$r^2_{10}  $\}& $0.027$&$0.010$&$0.013$&  $0.038$&$0.014$&$0.019$&  $0.044$&$ 0.020$&$ 0.030 $   \\
\hline
Im\{$r^2_{1-1} $\}& $-0.329$&$0.016$&$0.025$& $-0.319$&$0.020$&$0.027$& $-0.285$&$ 0.030$&$ 0.030 $  \\
\hline
$r^5_{11}    $&   $0.001$&$0.006$&$0.008$&  $0.012$&$0.008$&$0.004$&  $0.021$&$ 0.010$&$ 0.014 $   \\
\hline
$r^5_{00}    $&   $0.026$&$0.012$&$0.025$&  $0.070$&$0.020$&$0.009$&  $0.117$&$ 0.027$&$ 0.050 $  \\
\hline
Re\{$r^5_{10}$\} &  $0.137$&$0.005$&$0.011$&  $0.152$&$0.007$&$0.012$&  $0.144$&$ 0.010$&$ 0.014 $  \\
\hline
$r^5_{1-1}   $&   $-0.002$&$0.008$&$0.008$& $-0.012$&$0.010$&$0.008$& $-0.018$&$ 0.014$&$ 0.010 $ \\
\hline
Im\{$r^6_{10}  $\}& $-0.144$&$0.005$&$0.009$& $-0.135$&$0.007$&$0.004$& $-0.121$&$ 0.011$&$ 0.022 $ \\
\hline
Im\{$r^6_{1-1} $\}& $0.012$&$0.008$&$0.008$&  $0.007$&$0.010$&$0.007$&  $0.009$&$ 0.015$&$ 0.014 $  \\
\hline 
\end{tabular}
\vspace{1cm}
\caption{
The 15 combinations of spin-density matrix elements $r^{04}_{ik}$, $r^{\alpha}_{ik}$,
as obtained from the BPC data in three $t$ intervals. Statistical
and systematic uncertainties are given separately.
The data cover the kinematic range
$0.25<Q^2<0.85$~GeV$^2$, $20<W<90$~GeV, $0.6<M_{\pi\pi}<1.0$~GeV and 
$|t|<0.6$~GeV$^2$. The average $|t|$ values for the three intervals are 
$\langle |t|\rangle=0.04$~GeV$^2$, $\langle |t| \rangle=0.16$~GeV$^2$ and
$\langle |t|\rangle=0.37$~GeV$^2$, respectively.
}
\label{bpc_t}
\end{center}
\end{table}
%
%
%
%
\clearpage
\begin{table}[h]
%
\scalebox{0.6}[1.1]{
\begin{tabular}{|l|r|r|r|r|r|r|r|r|r|r|r|r|r|r|r|}
\hline
\multicolumn{16}{|c|}{ZEUS 1995 (BPC)}\\
\hline
&$r^{04}_{00}$&Re\{$r^{04}_{10}$\}&$r^{04}_{1-1}$&$r^1_{11}$&$r^1_{00}$&Re\{$r^1_{10}$\}&
$r^1_{1-1}$&Im\{$r^2_{10}$\}&Im\{$r^2_{1-1}$\}&$r^5_{11}$&$r^5_{00}$&Re\{$r^5_{10}$\}&
$r^5_{1-1}$&Im\{$r^6_{10}$\}&Im\{$r^6_{1-1}$\} \\
\hline
$r^{04}_{00}$& $1.000$ &&&&&&&&&&&&&& \\
\hline
Re\{$r^{04}_{10}$\}& $0.079$&$ 1.000$ &&&&&&&&&&&&& \\
\hline
$r^{04}_{1-1}$& $0.046$&$ 0.054$&$ 1.000$ &&&&&&&&&&&& \\
\hline
$r^1_{11}$& $0.101$&$ 0.011$&$ 0.584$&$ 1.000$ &&&&&&&&&&& \\
\hline
$r^1_{00}$& $-0.159$&$-0.025$&$-0.062$&$-0.414$&$ 1.000$ &&&&&&&&&& \\
\hline
Re\{$r^1_{10}$\}& $-0.068$&$-0.508$&$-0.127$&$-0.039$&$ 0.043$&$ 1.000$ &&&&&&&&& \\
\hline
$r^1_{1-1}$& $-0.327$&$-0.201$&$-0.082$&$-0.082$&$ 0.052$&$ 0.116$&$ 1.000$&&&&&&&& \\
\hline
Im\{$r^2_{10}$\}& $0.105$&$ 0.445$&$ 0.075$&$ 0.062$&$-0.039$&$-0.293$&$-0.095$&$ 1.000$&&&&&&& \\
\hline
Im\{$r^2_{1-1}$\}& $0.312$&$ 0.097$&$ 0.019$&$ 0.009$&$-0.029$&$-0.069$&$ 0.073$&$ 0.072$&$ 1.000$&&&&&& \\
\hline
$r^5_{11}$& $-0.095$&$ 0.231$&$ 0.014$&$-0.075$&$ 0.063$&$-0.038$&$ 0.014$&$ 0.011$&$ 0.050$&$ 1.000 $&&&&& \\
\hline
$r^5_{00}$& $0.248$&$ 0.578$&$ 0.053$&$ 0.087$&$-0.223$&$-0.452$&$-0.150$&$ 0.544$&$ 0.133$&$-0.300$&$ 1.000$&&&& \\
\hline
Re\{$r^5_{10}$\}& $0.299$&$ 0.080$&$-0.225$&$-0.138$&$-0.184$&$-0.201$&$-0.098$&$-0.084$&$ 0.023$&$-0.004$&$-0.044$&$ 1.000$ &&& \\
\hline
$r^5_{1-1}$& $0.003$&$-0.325$&$ 0.052$&$ 0.058$&$ 0.007$&$ 0.117$&$-0.015$&$-0.012$&$-0.068$&$-0.479$&$-0.051$&$ 0.078$&$ 1.000$&& \\
\hline
Im\{$r^6_{10}$\}& $-0.308$&$ 0.051$&$-0.359$&$-0.221$&$-0.356$&$ 0.031$&$ 0.009$&$-0.056$&$-0.057$&$ 0.044$&$ 0.034$&$ 0.056$&$-0.054$&$ 1.000$ & \\
\hline
Im\{$r^6_{1-1}$\}& $-0.009$&$ 0.337$&$-0.046$&$-0.074$&$ 0.014$&$ 0.062$&$-0.072$&$ 0.169$&$ 0.035$&$ 0.443$&$ 0.041$&$-0.040$&$-0.251$&$ 0.055$&$ 1.000$ \\
\hline 
\end{tabular}
}
%
\vspace{1cm}
\caption{Correlation matrix for the 15 combinations of spin-density 
matrix elements $r^{04}_{ik}$, $r^{\alpha}_{ik}$, as obtained from the 
BPC data. 
}
\label{tab2}
\end{table}
%
%
%
%
\clearpage
\begin{table}[h]
\scalebox{0.6}[1.1]{
\begin{tabular}{|l|r|r|r|r|r|r|r|r|r|r|r|r|r|r|r|}
\hline
\multicolumn{16}{|c|}{ZEUS 1995 (DIS)} \\
\hline
 &$r^{04}_{00}$&Re\{$r^{04}_{10}$\}&$r^{04}_{1-1}$&$r^1_{11}$&$r^1_{00}$&Re\{$r^1_{10}$\}&
$r^1_{1-1}$&Im\{$r^2_{10}$\}&Im\{$r^2_{1-1}$\}&$r^5_{11}$&$r^5_{00}$&Re\{$r^5_{10}$\}&
$r^5_{1-1}$&Im\{$r^6_{10}$\}&Im\{$r^6_{1-1}$\} \\
\hline
$r^{04}_{00}$& $1.000$ &&&&&&&&&&&&&& \\
\hline
Re\{$r^{04}_{10}$\}& $-0.014$&$ 1.000$ &&&&&&&&&&&&& \\
\hline
$r^{04}_{1-1}$& $0.020$&$-0.081$&$ 1.000$ &&&&&&&&&&&& \\
\hline
$r^1_{11}$& $0.061$&$-0.097$&$0.273$&$ 1.000$ &&&&&&&&&&& \\
\hline
$r^1_{00}$& $-0.011$&$-0.042$&$0.070$&$-0.381$&$ 1.000$ &&&&&&&&&& \\
\hline
Re\{$r^1_{10}$\}& $0.086$&$-0.219$&$0.146$&$-0.056$&$-0.086$&$ 1.000$ &&&&&&&&& \\
\hline
$r^1_{1-1}$& $-0.203$&$0.059$&$-0.060$&$-0.050$&$0.023$&$0.196$&$ 1.000$&&&&&&&& \\
\hline
Im\{$r^2_{10}$\}& $0.017$&$0.176$&$0.013$&$-0.032$&$-0.061$&$0.152$&$0.099$&$ 1.000$&&&&&&& \\
\hline
Im\{$r^2_{1-1}$\}& $0.182$&$-0.008$&$-0.028$&$0.041$&$0.005$&$0.170$&$0.433$&$0.095$&$ 1.000$&&&&&& \\
\hline
$r^5_{11}$& $-0.257$&$0.292$&$-0.050$&$-0.174$&$0.088$&$-0.038$&$-0.054$&$0.166$&$-0.086$&$ 1.000 $&&&&& \\
\hline
$r^5_{00}$& $0.172$&$0.422$&$-0.027$&$0.067$&$-0.221$&$-0.333$&$-0.004$&$0.254$&$0.042$&$-0.279$&$ 1.000$&&&& \\
\hline
Re\{$r^5_{10}$\}& $-0.131$&$0.331$&$-0.460$&$-0.170$&$-0.322$&$-0.221$&$0.023$&$-0.029$&$ -0.162$&$0.054$&$-0.004$&$ 1.000$ &&& \\
\hline
$r^5_{1-1}$& $0.015$&$-0.270$&$0.309$&$-0.022$&$-0.013$&$0.179$&$-0.107$&$0.195$&$-0.094$&$-0.156$&$-0.028$&$-0.038$&$ 1.000$&& \\
\hline
Im\{$r^6_{10}$\}& $0.124$&$0.076$&$-0.432$&$-0.220$&$-0.339$&$-0.134$&$-0.143$&$-0.214$&$0.102$&$0.021$&$0.092$&$0.506$&$-0.014$&$ 1.000$ & \\
\hline
Im\{$r^6_{1-1}$\}& $-0.071$&$0.244$&$0.109$&$-0.043$&$0.007$&$0.205$&$-0.045$&$0.284$&$-0.245$&$0.221$&$-0.002$&$0.001$&$0.286$&$0.013$&$ 1.000$ \\
\hline 
\end{tabular}
}
\vspace{1cm}
\caption{Correlation matrix for the 15 combinations of spin-density 
matrix elements $r^{04}_{ik}$, $r^{\alpha}_{ik}$, as obtained from the DIS data. 
}
\label{tab3}
\end{table}

%
%
\clearpage
\begin{table}
\renewcommand{\arraystretch}{1.3}
\def\largestrut{\vrule width 0pt height 30pt depth 20pt\relax}
\begin{center}
\begin{tabular}{|c|r@{$\ \pm\ $}l@{$\ \pm\ $}l|r@{$\ \pm\ $}l@{$\ \pm\ $}l|}
\hline
 \multicolumn{7}{|c|}{ZEUS 1995 (BPC) }  \\ 
\hline
& \multicolumn{3}{c|}{$0.6<M_{\pi\pi}<0.77$ GeV } 
& \multicolumn{3}{c|}{$0.77<M_{\pi\pi}<1.0$ GeV }   \\ 
\hline
$r^{04}_{00}$&     $0.311$&$ 0.014$&$ 0.013 $ &  $0.227$&$ 0.018$&$ 0.016 $ \\
\hline
Re\{$r^{04}_{10}$\}& $0.026$&$ 0.009$&$ 0.010 $ &  $0.032$&$ 0.010$&$ 0.005 $ \\
\hline
$r^{04}_{1-1}  $&  $-0.031$&$ 0.011$&$ 0.006$ & $-0.062$&$ 0.014$&$ 0.010 $ \\
\hline
$r^1_{11}    $&    $-0.030$&$ 0.012$&$ 0.009$ & $-0.054$&$ 0.015$&$ 0.007 $ \\
\hline
$r^1_{00}    $&    $0.020$&$ 0.019$&$ 0.042 $ & $-0.011$&$ 0.026$&$ 0.041 $ \\
\hline
Re\{$r^1_{10}$\}&    $-0.007$&$ 0.010$&$ 0.010$ & $-0.029$&$ 0.013$&$ 0.010 $\\ 
\hline
$r^1_{1-1}   $&    $0.321$&$ 0.014$&$ 0.016$  & $0.369$&$ 0.018$&$ 0.027 $  \\
\hline
Im\{$r^2_{10}  $\}&  $0.023$&$ 0.010$&$ 0.010$ &  $0.064$&$ 0.012$&$ 0.017 $  \\
\hline
Im\{$r^2_{1-1} $\}&  $-0.302$&$ 0.014$&$ 0.029$ & $-0.362$&$ 0.019$&$ 0.016 $ \\
\hline
$r^5_{11}    $&    $0.015$&$ 0.006$&$ 0.009 $  & $-0.002$&$ 0.007$&$ 0.006$ \\
\hline
$r^5_{00}    $&    $0.026$&$ 0.013$&$ 0.019$  & $0.077$&$ 0.015$&$ 0.017 $ \\
\hline
Re\{$r^5_{10}$\}  &  $0.152$&$ 0.005$&$ 0.008 $   &$0.129$&$ 0.007$&$ 0.012 $ \\
\hline
$r^5_{1-1}   $&    $-0.010$&$ 0.007$&$ 0.007$  &$-0.009$&$ 0.009$&$ 0.013 $ \\
\hline
Im\{$r^6_{10}  $\}&  $-0.155$&$ 0.004$&$ 0.004$  &$-0.125$&$ 0.006$&$ 0.006 $ \\
\hline
Im\{$r^6_{1-1} $\}&  $0.019$&$ 0.007$&$ 0.006$  &$0.007$&$ 0.009$&$ 0.008 $  \\
\hline 
\end{tabular}
\vspace{1cm}
\caption{The 15 combinations of spin-density matrix elements $r^{04}_{ik}$, $r^{\alpha}_{ik}$, 
as obtained from the BPC data in two $M_{\pi\pi}$ intervals. Statistical
and systematic uncertainties are given separately. 
The data cover the kinematic range
$0.25<Q^2<0.85$~GeV$^2$, $20<W<90$~GeV, $0.6<M_{\pi\pi}<1.0$~GeV and 
$|t|<0.6$~GeV$^2$. 
}
\label{bpcvsmass}
\end{center}
\end{table}

\clearpage

\clearpage

\clearpage

\clearpage

\clearpage

\clearpage

\clearpage

\clearpage


\begin{thebibliography}{20}
  

\bibitem{jim} 
J.A.~Crittenden, ``Exclusive Production of Neutral Vector 
Mesons at the Electron-Proton Collider HERA", {\rm Springer Tracts in Modern
Physics}, Volume~140, Springer, Berlin and Heidelberg (1997),
and references therein.

\bibitem{sakurai} 
J.J.~Sakurai, ``Currents and Mesons", University of Chicago Press, Chicago
  (1969);\\
H.~Fraas and D.~Schildknecht, Nucl. Phys. B14 (1969)~543.

\bibitem{bauer} 
T.H.~Bauer et al., Rev. Mod. Phys.~50 (1978)~261, Erratum ibid. 51 (1979) 
407, and references therein.

\bibitem{collins}
P.D.B. Collins, ``An Introduction to Regge Theory and High Energy
Physics", Cambridge University Press, Cambridge~(1977).

\bibitem{ballam} J.~Ballam et al., Phys. Rev. D7 (1973)~3150.


\bibitem{joos} P.~Joos  et al., Nucl. Phys.  B113  (1976)~53.

\bibitem{chio} CHIO Collab., W.D. Shambroom et al., Phys. Rev. D26 (1982) 1.

\bibitem{psi} H1 Collab., S. Aid et al., Nucl. Phys. B472 (1996) 3;\\
ZEUS Collab., J. Breitweg et al., Z. Phys. C75 (1997) 215.

\bibitem{rhodis95} ZEUS Collab., J. Breitweg et al.,
Eur. Phys. J. C6 (1999) 603.

\bibitem{h1} H1 Collab., C. Adloff et al., DESY Report DESY 99-010, hep-ex/9902019 (1999), 
to appear in Eur. Phys. J. C.

\bibitem{zeusrho93}ZEUS Collab., M. Derrick et al.,
Z. Phys. C69 (1995) 39.

\bibitem{h1rho93} H1 Collab., S. Aid et al.,
Nucl. Phys. B463 (1996) 3.

\bibitem{zeusrho94}ZEUS Collab., J. Breitweg et al.,
Eur. Phys. J. C2 (1998) 247.

\bibitem{workshop96} 
H. Abramowicz et al., in {\em Proceedings of the Workshop on Future Physics at 
HERA, Volume~II},
edited by G.~Ingelman, A. De Roeck and R. Klanner (DESY, Hamburg, Germany, 
1996), p.~635, and references therein;\\
W. Koepf et al., ibid., p.~674, and references therein.

\bibitem{cfs}
J.C. Collins, L.L. Frankfurt and M. Strikman, Phys. Rev. D56 (1997) 2982.

\bibitem{zeusrhodis93}ZEUS Collab., M. Derrick et al.,
Phys. Lett. B356 (1995) 601.

\bibitem{h1rhodis93} H1 Collab., S. Aid et al.,
Nucl. Phys. B468 (1996) 3.

\bibitem{h1rhodiss94} H1 Collab., C. Adloff et al.,
Z. Phys. C75 (1997) 607.


\bibitem{ivanov} D. Yu. Ivanov and R. Kirschner, Phys. Rev. D58 (1998) 114026.



\bibitem{kolya} E.V. Kuraev, N.N. Nikolaev and B.G. Zakharov, JETP Lett.
68 (1998) 696 and Pisma Zh. Eksp. Teor. Fiz. 68 (1998) 667;\\
N.N. Nikolaev, Proceedings of the DIS99 workshop, April 19-23, 1999, Zeuthen, 
Germany, Eds. J. Bl\"umlein and T. Riemann, to appear in Nucl. Phys. B 
(Proc. Suppl.);\\
I. Akushevich, I. Ivanov, N.N. Nikolaev and A. Pronyaev, private communication.

\bibitem{cudell} I. Royen, Proceedings of the DIS99 workshop, 
April 19-23, 1999, Zeuthen, Germany,
Eds. J. Bl\"umlein and T. Riemann, to appear in Nucl. Phys. B (Proc. Suppl.)
and Li\`ege University preprint ULG-PNT-99-1-IR;\\
I. Royen, private communication.


\bibitem{diehl} M. Diehl et al., Phys. Rev. D59 (1999) 34023.

\bibitem{schildknecht} D. Schildknecht, G. A. Schuler and B. Surrow, 
Phys. Lett. B449 (1999) 328.

\bibitem{zeus} 
ZEUS Collab., ``The ZEUS Detector Status report", DESY (1993).

\bibitem{rhodis94} ZEUS Collab., M. Derrick et al., Phys. Lett. 
B356 (1995) 601.

\bibitem{ref:angle} 
K.~Schilling and G.~Wolf, Nucl. Phys. B61 (1973)~381.

\bibitem{Heiko} 
H.~Beier, Ph.D. thesis, Hamburg University (1997), DESY Internal Report 
DESY F35D-97-06.

\bibitem{Teresa} 
T. Monteiro, Ph.D. thesis, Hamburg University (1998), DESY Internal Report 
DESY-THESIS-1998-027.

\bibitem{ref:jetset} 
T.~Sj\"ostrand, Comp. Phys. Commun. 39 (1986)~347;\\
T.~Sj\"ostrand and M. Bengtsson, Comp. Phys. Commun. 43 (1987)~367.

\bibitem{ref:muchor} 
K.~Muchorowski, Ph.D. thesis, Warsaw University (1998).

\bibitem{ref:heracles} 
A.~Kwiatkowski, H.~Spiesberger and H.-J.~Moehring,
in {\em Proceedings of the Workshop on Physics at HERA, Volume~III},
edited by W.~Buchm\"uller and G.~Ingelman (DESY, Hamburg, Germany, 1991),
p.~1294.

\bibitem{soeding} P.~S\"{o}ding, Phys. Lett. 19 (1966)~702.

\bibitem{ryskin} M.G. Ryskin and Yu.M. Shabelski, Phys. At. Nucl. 61 (1988) 81.


\bibitem{reviews} 
G. Alberi and G. Goggi, Phys. Rep. 74 (1981) 1;\\
K. Goulianos, Phys. Rep. 101 (1983) 169;\\
N.P. Zotov and V.A. Tsarev, Sov. Phys. Uspekhi  31 (1988) 119;\\
G. Giacomelli, Int. J. Mod. Phys. A5 (1990) 223.


\end{thebibliography}
\end{document}